# Title

# Complexity as Axiomatic Incompleteness

Zhifeng Ma, Zhangang Han*

School of Systems Science, Beijing Normal University, Beijing 100875, People's Republic of China

*Corresponding author Zhangang Han.

**Email:** zhan@bnu.edu.cn

**Tel:** +86 13488806949

**Author Contributions:** Z.H. conceived the idea and designed the research. Z.M. carried out the implementation of the proof. Z.H. supervised the research. Z.M and Z.H. wrote the manuscript.

**This file includes:**

Main Text
Figures 1 to 3

**Abstract**

Information theory has long sought to uncover the universal laws governing complex systems. Yet information is not merely quantified; it is also interpreted by observers. We formalize this cognitive aspect through a Generalized Incompleteness Theory (GIC), which integrates axiomatic logic, Probably Approximately Correct (PAC) learning, and Gödel's incompleteness theorem to establish a connection between complexity and formal incompleteness. We introduce the notions of interpreter-based information and PAC complexity, demonstrating that certain outputs can be PAC-learnable even when their corresponding interpreters cannot be constructed. We further show that every nontrivial formal system contains a PAC-learnable incompleteness space of dimension (2^n-n-1). The GIC framework unifies several foundational results—including Cantor's diagonal argument, Gödel's incompleteness theorem, and the Turing halting problem—within a categorical and topological framework. We further show that physical models, including four-dimensional spacetime in relativity and the 11- and 26-dimensional spaces arising in string theory, admit natural correspondences with incompleteness spaces, while critical states can be understood in terms of symmetry-breaking selections of axioms. Our framework identifies incompleteness as a universal feature underlying complexity, criticality, and observability, thereby revealing fundamental limits on description and cognition in any axiomatizable theory.

## Introduction

Complex system is involved in a board range of research fields whose components are nonlinearly interacting and exhibit the property that the whole is not a simple sum of its parts[1,2,3,4,5,6,7,8,9]. Complexity has not yet had a unified and consensus definition, due to the wide range of applications across space and time scales[10,11,12,13,14,15,16,17,18,19,20,21].

Information is usually perceived as a measure of uncertainty eliminated, hence entropy as a measurement of complexity. Original information theory is designed for objective binary information. The cognitive process of information, however, requires and the information receiver's interpretation of the information. The interpretation of information could be regulated by Probably Approximately Correct (PAC) learnable theory to find the best interpreter in polynomial time steps[22,23].

To formalize the process of interpreting information, the Gödel incompleteness theorems form the basis of the analysis and establish the limits of the cognitive process [24,25,26,27,28,29,30]. This study: (a) establishes the PAC-complexity framework by defining the system, interpreter-based information, and complexity;(b) constructs the Gödel incompleteness space to cover all PAC-learnable outputs for which the corresponding interpreter cannot be found; (c) employs wave function to describe distribution to which quantum mechanics can be easily linked.

This study presents the general Gödel incompleteness theorem (GIC) and provides a calculation of its dimensions. These results are then used to:(a) demonstrate the Cantor diagonalization argument, the original Gödel incompleteness theorem and the Turing halting problem within the same framework;(b) provide a mathematical description of the critical state as a Gödel incompleteness space, which can be applied to a variety of studies;(c) correspond the dimensions of incompleteness to the dimensions of fundamental physics.

## Materials and Methods

### General Incompleteness Space

#### *General Gödel incompleteness theorem*

Figure 3 illustrates the cognitive processes and main proof category diagrams in this section.

For $I$ is the information before interpretation. $H$ is the interpreter space that contains all the possible interpreters. $S$ is the space of cognition output. $\psi$ is a wave function to describe the distribution of $S$. We can capture the cognition process as follows,

$$HI = S = \lambda|\psi|^2 \tag{1}$$

Dirac notation form:

$$H|I\rangle = |S\rangle = \lambda \sum_i \langle e_i|\psi\rangle \langle\psi|e_i\rangle|e_i\rangle \tag{2}$$

Let $S$ be formal system to describe its inner logic procedure. We select different interpreters to form the interpreter space $H$ and give a distribution $\lambda|\psi|^2$ to each object in $S$ to form the formal system $S$ into a space. Proper operations are defined to make $H$ and $S$ linear spaces and use $H$ and $S$ to form tensor spaces $H_\otimes$ and $S_\otimes$ respectively whose object is the subspace of $H$ or $S$ (see details in section 3 [31]). Therefore, we could get the GIC as follows:

If H is non-trivial, there exists some object $\boldsymbol{S_{Gödel}}$ in $S_\otimes$ that is PAC learnable but does not have the corresponding object in $H_\otimes$ so that $HI = \boldsymbol{S_{Gödel}}$.

If the number of axioms of $H$ is greater than 2, i.e., $rank(H) \geq 2$, we call H non-trivial system.

#### *PAC-complexity framework*

We define a system S as a category with objects, morphisms and its environment. Environment is the automorphism of $S$ indicating the basic elements that form the objects and the basic symmetries on S. Given distribution $s(x)$ interpreted object $I$ and interpreter $h$, then we give the information amount $\mathcal{I}$ and its complexity measurement $c$ as follows (see details in section 1 [31]),

$$\mathcal{I} = \int h(I)(s(x))dx \tag{3}$$

$$c(h) = \int s(x) \cdot [h(I)]\big(s(x)\big)dx \tag{4}$$

where $h(I)$ is an operator as a whole. On the comparison with the original PAC learnable theory, we substitute finding proper algorithms into finding proper interpreters and minimizing the loss function into matching the minimum loss complexity $\left|\widehat{c(h^*)} - c(h^*)\right|$. Given any interpreted distribution, any complexity accuracy $\epsilon$, and any incorrect possibility $\delta$, the best interpreter $h^*$ could be found in polynomial time steps as long as the size of the training set $m$ is greater than a certain number $N = \frac{ln\frac{2|H|}{\delta}}{2\epsilon^2}$ [32,33]. Hence, we establish the PAC-complexity framework as above (see details in section 2 [31]).

***Incompleteness space***

Here we are going to prove that there exists some object Incomplete space $\boldsymbol{S_{Gödel}}$ in $S_{\otimes}$ that does not have the corresponding object in $H_{\otimes}$ so that $HI = \boldsymbol{S_{Gödel}}$ (see details in section 4 [31]).

The theorem "Incompleteness of interpreter space for proposition space" which is in line with the GIC goes as following:

If H is non-trivial, then such bijection $H_{\otimes} \leftrightarrow S_{\otimes}$ doesn't exist.

Consider the sentence that caused incompleteness:

$$\boldsymbol{S_{Gödel}} = S_B \otimes S_A + S_{\neg B} \otimes S_{\neg A} \quad (5)$$

$n = \dim(H)$ means H has at most $n$ linearly independent eigen vectors, which means we can compose an entangled $\boldsymbol{S_{Gödel}}$ space by choosing k of n eigen vectors, so the same k eigen vectors could compose linear independent entangled eigen vectors of number $\binom{n}{k}$. We get

$$\dim(\boldsymbol{S_{Gödel}}) = 2^n - \binom{n}{1} - \binom{n}{0} = 2^n - n - 1 \quad (6)$$

**Results**

**Mathematical Logic and Quantum Entanglement**

***Turing halting problem***

Given the definitions of the General Incompleteness Theorem and the dimension of the incompleteness space provided in the 'Materials and Methods (MM)' section, we can naturally apply the basic Gödel incompleteness sentence structure, $\boldsymbol{S_{Gödel}} = S_B \otimes S_A + S_{\neg B} \otimes S_{\neg A}$ , to derive the original Gödel incompleteness theorems and the proofs of the Turing halting problem, as demonstrated in the Supporting Information [31], sections 5.2 and 5.3, and schematically visualized in Fig. 1.

***Cantor diagonal proof***

Here we re-prove that the set of real numbers is larger than the set of natural numbers [34,35] , using the GIC (see section 5.1 of [31]).

Let $n \to \infty$, especially, let $n = \aleph_0$, then $\dim(S_{Gödel}) = 2^{\aleph_0} = \aleph_1$.

One of the greatest achievements of this formula is that it provides a means of discussing the continuum hypothesis. It demonstrates that a denumerable infinity $\aleph_0$ could construct $\aleph_1$ by allowing for any type of limit. Similar to the $s_{\text{Gödel}}$ constructed in MM section, we provide any real number a between 0 and 1 with an infinite decomposition:

$$\mathrm{a} = 0 + \mathrm{a}_1 \times 10^{-1} + \mathrm{a}_2 \times 10^{-2} + \mathrm{a}_3 \times 10^{-3} + \cdots \tag{7}$$

Consider a as an infinite sum of eigenvectors $10^{-\mathrm{i}}$, and let $\mathrm{a}_\mathrm{i} \in \{0,1,\dots,9\}$. Then any real number in this interval can be regarded as an expansion of the original rational numbers $10^{-\mathrm{i}}$ and $\{0,1,\dots,9\}$. Therefore, the $\aleph_1$ is the PAC learnable expanded space of $\aleph_0$ , while some of the numbers in $\aleph_1$ cannot be found in $\aleph_0$.

***Quantum Entanglement***

If we apply the GIC to quantum mechanics[36,37], we predict that certain states exist but cannot be observed directly. (See section 5.4 of [31] for details.)

Choose the spin of particle 1 as the axioms of B and the spin of particle 2 as the axioms of A. There is a $s_{Gödel}$ state which is entangled but cannot be measured. For one particle spinning in two different directions composing a Hilbert space $H$, one version of $S_{Gödel}$ could be written as follows:

$$S_{Gödel} = \uparrow\otimes\uparrow + \downarrow\otimes\downarrow \tag{8}$$

The corresponding $s_{Gödel}$ will destroy the completeness of $H$.

And the other $S_{Gödel}{}'$ can be written as follows:

$$S_{Gödel}{}' = \uparrow\otimes\downarrow + \downarrow\otimes\uparrow \tag{9}$$

The corresponding $s_{Gödel}{}'$ will indicate that the consistence of $H$ cannot be proved within $H$.

However, if $\dim(H) = 1$, all the above paradoxes will disappear, resulting in an observable eigenstate.

***Critical State as Incomplete Space***

Figure 2 provides a schematic representation of the central idea of this chapter: the correspondence between the dimensionality of the incompleteness space and the degrees of freedom of a theory.

***Syntax Duality and Semantic Duality***

Here we present a sufficient condition for applying GIC. (See section 6.1 of [31] for details.)

Let there be a pair of adjoint functors mapping between categories $X$ and $Y$ [38,39,40]. This is equivalent to the syntactic duality between $X$ and $Y$ within our framework, which means every sentence in $X$ or $Y$ could be translated into each other bit by bit. If category $X$ has a topos structure (i.e., it contains at least first-order predicate logic) and has a construction starting point (axiom) [41], while if there are n starting points, under the condition that free construction is allowed, $Y$ will have $2^n - n - 1$ orthogonal starting points of the incomplete theorem. This is equivalent to the incompleteness of the semantic duality, which means the logic procedure in $X$ cannot approach certain sentences in $Y$ that have been translated into $X$.

The requirement of the theorem for category $X$ can be relaxed to the extent that $X$ can be described in the language of set theory by Yoneda embedding [41] to make $X$ a topos.

***Quantum observability***

We now prove the following statement: The quantum eigenstates iff PAC learnable iff observable, while the quantum superposition states iff PAC unlearnable iff unobservable [42]. (See section 6.4 of [31] for details.)

Understanding the family of the full trajectory of an object's movement requires additional dimensions. All trajectory families are combined to form a state, which behaves like a critical state that is ready to collapse into an eigenstate due to observation.

In the case of double-slit interference and without observation, the superposition state represents an incomplete space spanned by various possible orthogonal base paths. Therefore, before path observation, the pattern on the screen displayed the particle positions in this "completely" incomplete space. After observation of the paths, only those through a single slit are shown, as these are the only PAC-learnable states.

***Phase transition***

For a system in which the automorphism symmetry group of each object contains the fundamental group $\mathbb{Z}^2$ [43] (i.e. for any object within the system, there are two non-equivalent construction loops that can return to themselves), if the syntactic translation of another system is complete, then its semantic logical translation (in the sense of allowing "limits") is incomplete. This proposition is a sufficient condition for the dimensional version of the GIC. This is because, if each propositional object can return to itself through a construction loop of some kind, then this loop corresponds to a construction axiom. If each propositional object has at least two non-equivalent loops, then the system contains at least two non-equivalent construction axioms. Therefore, we can formally transform the study of a system's logical structure into the study of the topological space in which its logic is located.

As the syntax and semantics duality matching appropriately, the critical state is the incompleteness space. The number of phases is the number of dimensions of the base space $n$. The number of order parameters required to completely describe the critical state is equal to the number of dimensions of the incomplete space $2^n - n - 1$.

For example, the Lotka–Volterra model [44] has 3 terminal states, while the complete description of the critical state of this model requires $4 = 2^3 - 3 - 1$ parameters (see section 6.5 of [31]).

Symmetry breaking causes the critical state to fall into terminal states, the number of which corresponds to the axiom number of the base space. Therefore, the number of types of symmetry breaking is equivalent to the dimension of the base space.

***Zero absolute temperature***

If the phases (such as all the spins being either up or down in a ferromagnet) are taken as the base space, the incompleteness space as a whole can be understood in terms of several parameters of incomplete space dimensions (such as temperature). However, if we consider the ferromagnetic particles individually, the computational complexity of the number of parameters required to describe the incomplete space formed by the entanglement of n particles is $O(2^n)$. This is a PAC unlearnable problem; that is to say, this transition process cannot be observed.

In other words, throughout the entire process of achieving absolute zero[*45*], it is necessary to conduct PAC-learnable observations on all possible entangled and incomplete states that arise. Such an observation process is bound to consume energy in the sense of information theory, forming an exponential potential barrier when approaching absolute zero (see section 6.6 [31]).

***Dimensions of Fundamental Physics Frameworks***

We demonstrate here that the dimension of the incompleteness space is equivalent to the dimensions of fundamental physics frameworks. (see section 6.3 [31])

***Electromagnetism***

$n = 2$， $\dim(\boldsymbol{S}_{\boldsymbol{Gödel}}) = 1$

Correspondingly, given electric field E and magnetic field B as two axioms, there will be one certain type of incompleteness axiom

$$S = E \otimes B + \neg E \otimes \neg B \qquad (10)$$

This corresponds to the synchronization of the electric and magnetic fields of electromagnetic waves simultaneously, predicting the characteristics of the quantum mechanical wave-particle duality and the anti-Bell inequality of electromagnetic waves.

The notation $\neg$ represents rotation by 90 degrees, so that $\neg E\ and \neg B$ are orthogonal to $E\ and\ B$, respectively.

***Relativity***

$n = 3$， $\dim(\boldsymbol{S}_{\boldsymbol{Gödel}}) = 4$

For the Lorentz transformation group SO (1,3) [46], it can be divided into the direct sum of two parts SU (2). The fundamental group of SO (1,3) is $\mathbb{Z} / 2\mathbb{Z} = \{\pm 1\}$, thus it contains two loops. Translation transformation contains a loop. Therefore, the Poincaré transformation group as a whole is equivalent to three loops - three axioms. Therefore, the incomplete space formed by its entanglement has four dimensions.

***Superstring theory***

$n = 4$, $\dim(\boldsymbol{S}_{\boldsymbol{Gödel}}) = 11$

From the perspective of symmetries [47], the Poincaré transformation group plus $U(1) \times SU\ (2) \times SU\ (3)$ is the symmetry of the four fundamental forces. Among $U(1) \times SU\ (2) \times SU\ (3)$, only the fundamental group $U(1)$ has 1-dimension non-trivial fundamental group $\mathbb{Z}$, so the sum is a 4-dimension axiom set forming 11-dimension incompleteness space.

From the perspective of fundamental forces [48], it requires four kinds of symmetry breaking to collapse into four phases corresponding to four major types of gauge bosons, therefore, at least 11 dimensions are needed for complete description.

***Boson String theory***

$\boldsymbol{n = 5, \ dim(S_{Gödel}) = 26}$

The Poincaré invariance (3-dimension axioms) plus Diff invariance (1-dimension axiom) and Weyl invariance (1-dimension axiom) [49] capture the symmetry of boson string theory. Therefore, this theory's entire incompleteness space is 26 dimensions, which corresponds to the 26-dimension space in boson string theory.

## Conclusion

This study presents the PAC-complexity framework, which is a mathematical model of cognitive information processing. Using this framework, we prove the GIC and provide a calculation of the dimension of the incompleteness space. The framework combines a subjective understanding of information complexity with an objective understanding. The GIC shows that, while some information carriers are PAC-learnable, the corresponding information interpreter cannot be found.

Four elementary mathematical theorems are proven to validate the GIC and demonstrate its potential for generalization. This potential is further illustrated by presenting a sufficient application condition and a duality of syntax and semantics. Furthermore, this duality is developed into symmetry or fundamental group automorphisms of the category. Remarkably, the logical study of a theoretical framework has evolved into the study of the topological structure of the system.

The critical state, which has been widely studied in physical and non-physical systems such as the prey-predator model, is given a mathematical formalization within this framework. It transpires that the critical state is an incompleteness space, and the terminal state, along with its phase transition or symmetry breaking, is an axiom of the base space.

The dimension correspondences of fundamental physics theories are presented, implying that phenomena predicted by incompleteness theory are logically true, though they cannot be proven directly by experimental observation.

This study draws the limits of scientific description capability, which is the incompleteness space for any theory that can be formalized with axiom system.

## Acknowledgments

This work was supported by National Natural Science Foundation of China [62176022].

The authors would like to thank Tianyi Wu, Weizhen Tang, Yifan Zhang, Peizhi Yu, Zhen Kang, Pengju Song, Guozheng Lin, Tingting Xue, Huaiyu Wang, and Zhongjie Zhang from BNU School of Systems Science for fruitful discussions. Thanks to Prof. Minyong Guo from BNU School of Physics and Astronomy for discussion on string theory.

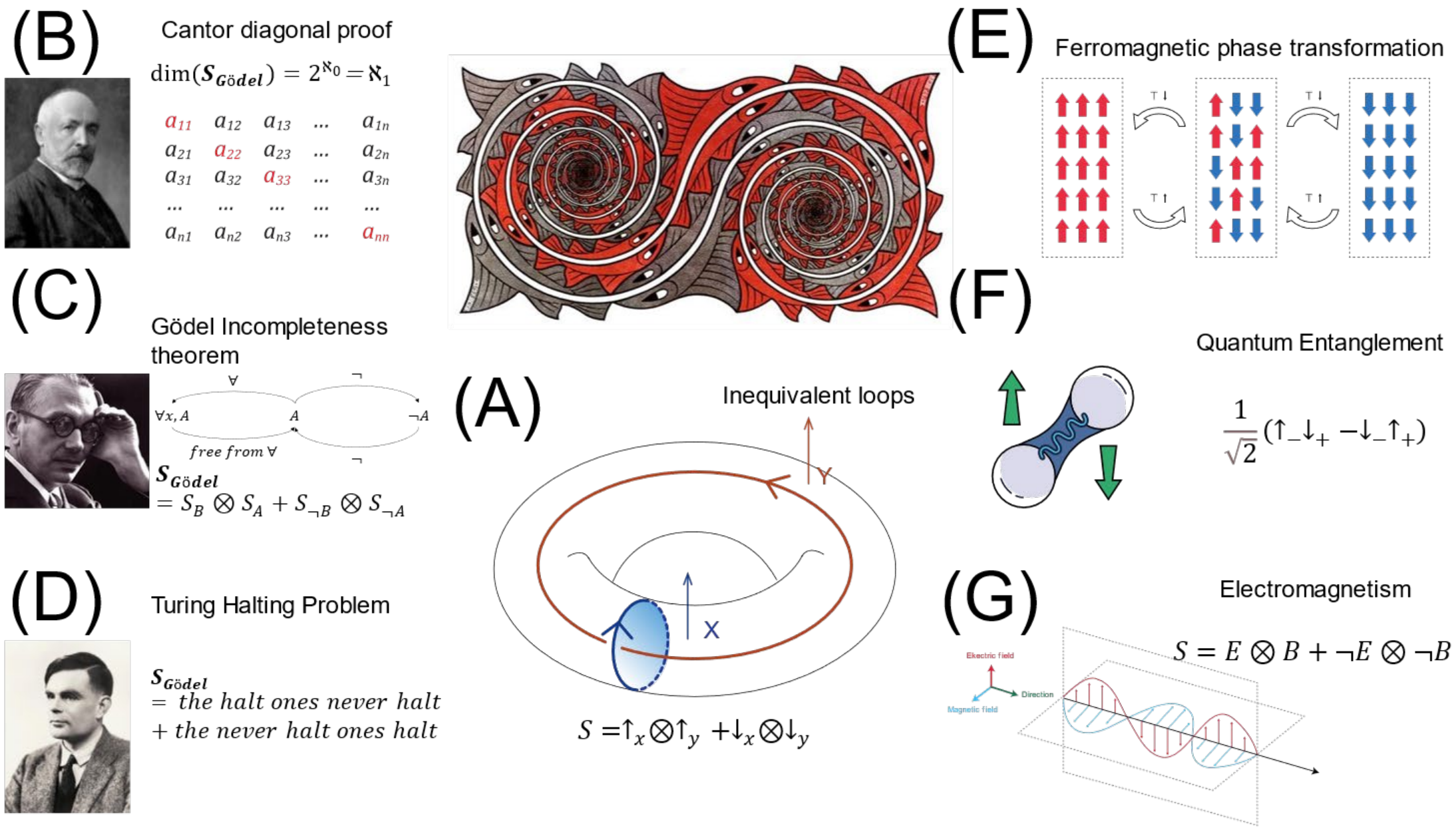

**Figure 1. The applications of the GIC.** (**A**) shows if a system has two inequivalent loops for each object in it, the incompleteness space could be constructed in an entangled way, which is topologically equivalent to put this system on a torus. The painting on the above of the torus is *Whirlpools* by Maurits Cornelis Escher in 1957 to demonstrate the entangled loops could span out an incompleteness space. (**B**) shows the real number set can be constructed from the natural number set by the GIC. (**C**) shows we can re-prove the original Gödel Incompleteness theorem using GIC. (**D**) shows we can re-prove the Turing halting problem using GIC. (**E**) shows we can describe critical state such as the ferromagnetic phase by the incompleteness space of GIC. (**F**) shows the quantum entanglement is strictly the incompleteness sentence in GIC. (**G**) shows we apply GIC on classical electromagnetism to directly get a quantum-behavior entity, i.e., the photon or the electromagnetic wave, that one cannot capture its entire properties by classical electromagnetism.

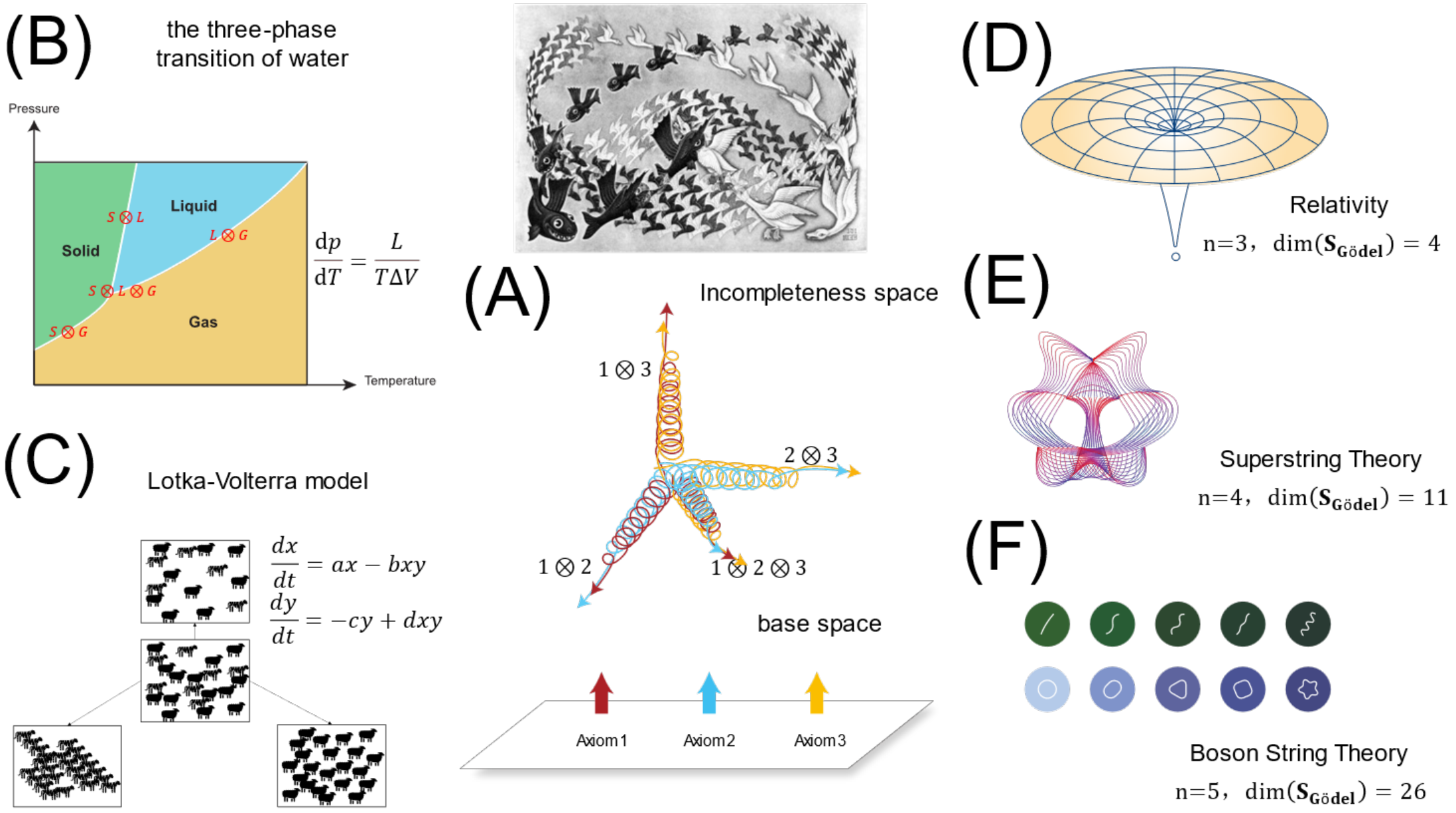

**Figure 2. The correspondence between the dimension of the incompleteness space and model degree of freedom or dimension of a theory.** (**A**) shows how an incompleteness space is spanned by entangling the axiom eigen vectors of the base space to create its basis. The painting on the above of the incompleteness space is *Predestination* by Maurits Cornelis Escher in 1951 to show the entangled axioms could span out an extremely complex incompleteness space. (**B**) shows three basic phases of water as axioms of the base space could form four-dimensional incompleteness space which need four order parameters to completely describe the phase transition critical state, as in the Clapeyron Equation. (**C**) shows that the Lotka-Volterra model, the prey-predator model, has the three terminal states as axioms of the base space so that the least complete description of the undetermined critical state in the middle would need four parameters in line with the dynamical equations of this model. (**D**), (**E**) and (**F**) suggest that the dimensional correspondence between the fundamental theories in physics and the incompleteness space is not accidental.

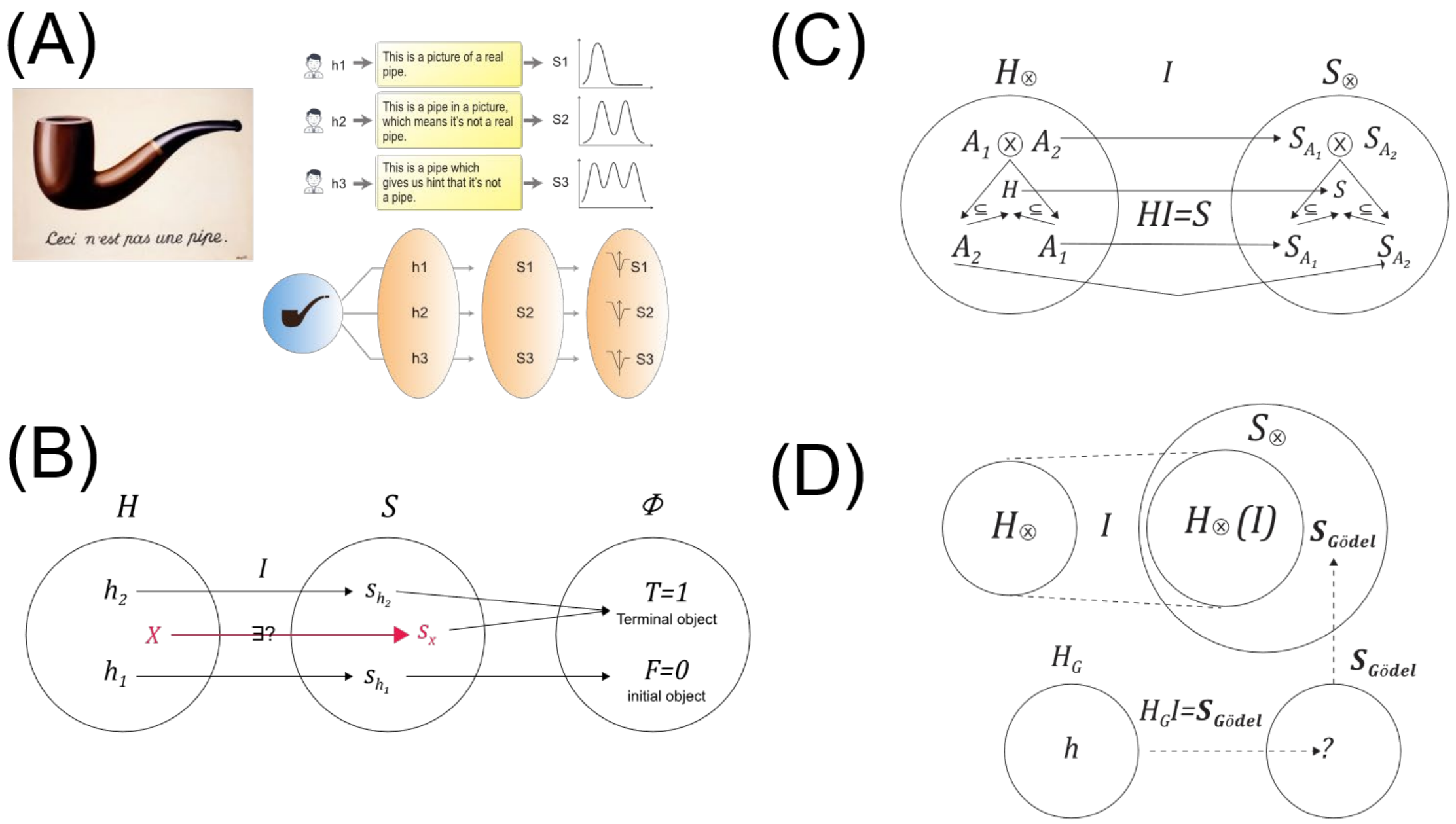

**Figure 3. The cognitive process using category language and the main category diagram.** (**A**) illustrates an example of cognizing the same painting (*The Treachery of Images-This is Not a Pipe* by Rene Magritte in 1929) from different interpreters $(h_1, h_2, h_3)$. Different interpreters receive different information, which corresponds to different information distribution $(s_1, s_2, s_3)$. (**B**) shows the category diagram of the cognitive process. (**C**) shows the category diagram of the definition of the tensor space $H_{\otimes}$ and $S_{\otimes}$. (**D**) shows the category diagram of the relationship between the tensor spaces and the incompleteness space $S_{Gödel}$ with its candidate interpreter $H_G$.

# **Supplement Material for**
# Complexity as Axiomatic Incompleteness

Zhifeng Ma, Zhangang Han*

School of Systems Science, Beijing Normal University, Beijing 100875, People's Republic of China

*Corresponding author Zhangang Han.

**Email:** zhan@bnu.edu.cn

**This file includes:**

Supplement Material text
Figures S1 to S12

**Supplement Material Text**

## *0 Introduction*

Complex system is involved in a board range of research fields whose components are nonlinearly interacting and exhibit the property that the whole is not a simple sum of its parts [1–9]. Complexity has not yet had a unified and consensus definition, due to the wide range of applications across space and time scales [10–21].

Information is usually perceived as a measure of uncertainty eliminated, hence entropy as a measurement of complexity. Original information theory is designed for objective binary information. The cognitive process of information, however, requires and the information receiver's interpretation of the information. The interpretation of information could be regulated by Probably Approximately Correct (PAC) learnable theory to find the best interpreter in polynomial time steps [22,23].

The scientific method is the portmanteau of instruments, formalisms, and experimental practices that succeed in discovering basic mechanisms despite the limitations of individual intelligence. There are, however, phenomena that are hidden in plain sight. These are the phenomena that we study as complex systems: the convoluted exhibitions of the adaptive world — from cells to societies.

Paradoxically, the complex world is one that we can, in many senses, perceive and measure directly. Unlike distant stars or nearby minerals that require a significant increase in optical capability to arrive at insights into their elementary properties, behavior — both individual and collective — seems to present itself in ways that can be investigated rather modestly through observation or experiment.

August 1972 saw the publication of Philip Anderson’s essay ‘More is different’ [2]. In it, he crystallized the idea of emergence, arguing that “at each level of complexity entirely new References properties appear” — that is, although, for example, chemistry is subject to the laws of physics, we cannot infer the field of chemistry from our knowledge of physics.

Complexity has always been one of the most fundamental subjects of information theory and complex system science. Despite numerous attempts to give a formal definition of complexity, our understanding of the complexity is not general and still far from being complete.

In computational learning theory, the theory of the learnable, proposed in 1984 by Leslie Valiant referred to as probably approximately correct (PAC) learning is a framework for mathematical analysis of machine learning. A class of concept is learnable if there exists an algorithm that can finish in polynomial time in the complexity of the concept and any arbitrary approximation ratio, probability of success on a given distribution of the samples. An important innovation of the PAC framework is the introduction of computational complexity theory concepts to machine learning. Nowadays, the mainly topic about PAC learning is about PAC-Bayesian generalization for like equivariant networks, offline contextual bandits, optimization and generalization in networks of neurons, etc.. The core idea of PAC-Bayesian generalization is to set the possibility distribution of the outcome into Bayesian possibility, which is conditional probability, not a priori probability. PAC-Bayesian learning of optimization algorithms is a significant development of such generalization of PAC learning[22,23].

In this paper, we will use this basic idea in the field of machine learning, PAC learning, to write the information processing procedure into mathematical form. The generalization function, the hypothesis, is considered as an interpreter and all the candidate hypotheses form the entire interpreter space. With the probability distribution in this idea, we can combine the information theory into this discussion and the generalization error would be the difference between the complexity under the candidate interpreter and the complexity that the objective information should have as it is.

Because our understanding of objectivity is limited, it is necessary to introduce PAC and Gödel's incompleteness theorem to understand the limit of objective facts.

To formalize the information interpretation process, Gödel incompleteness theorems is the backbone of analysis and lay the limit for where the cognition process can get. This study (a) defines the system, interpreter-based information, complexity and establishes the PAC-complexity framework; (b) constructs the Gödel incompleteness space to cover every interpreted output that is PAC learnable but the corresponding interpreter cannot be found; (d) employs wave function to describe distribution while quantum mechanics can be easily linked to.

Gödel's incompleteness theorems are two theorems of mathematical logic that are concerned with the limits of provability in formal axiomatic theories. These results, published by Kurt Gödel in 1931[24], are important both in mathematical logic and in the philosophy of mathematics. The theorems are widely, but not universally, interpreted as showing that Hilbert's program to find a complete and consistent set of axioms for all mathematics is impossible.

Gödel proves his first incompleteness theorem (G1) for a certain formal system P related to Russell–Whitehead's Principia Mathematica based on the simple theory of types over the natural number series and the Dedekind–Peano axioms. Gödel announces the second incompleteness theorem (G2) in an abstract published in October 1930: no consistency proof of systems such as Principia, Zermelo–Fraenkel set theory, or the systems investigated by Ackermann and von Neumann is possible by methods which can be formulated in these systems. Gödel sketches a proof of G2 and promises to provide full details in a subsequent publication. This promise is not fulfilled, and a detailed proof of G2 for first-order arithmetic only appears in a monograph by Hilbert and Bernays[24].

Our goal is to combine the fundamental theory of machine learning, information theory and mathematical logic together to achieve the proof of Gödel incompleteness theorem again but more generally. That means we could put several different topics under the same mathematical framework to discuss hitherto undone for the discussion of Gödel incompleteness theorem has been restricted to the pure mathematics or just mentioned their possible relationships with other fields briefly. The idea of introducing interpreters into the discussion of the Gödel incompleteness theorem could be traced back to representation system which uses representation theory on the formal systems to deduce the Gödel incompleteness theorem.

We can show that there are some objects that PAC can learn, but the best interpreter is not found among the alternatives. A strong enough machine algorithm cannot be interpretable in the face of any object. There are always objects that make a strong enough machine learning algorithm uninterpretable. Inspired by the traditional definition of complexity, this paper defines interpreter and then defines complexity under interpreter. Now there is the problem of choosing the most suitable interpreter, and heuristically finding the best interpreter is equivalent to using PAC to find the most suitable interpreter. We extend the scope of traditional complexity to include the influence of interpreters on complexity, and define interpreter-based complexity measures. For fixed cognitive individuals, there are different interpreters to recognize. Different interpreters will get different complexity, so fitting the most appropriate generalization complexity becomes the way to choose the most appropriate interpreter, which is consistent with the most basic theory of PAC learnable machine learning.

In the context of such a basic framework of machine learning intersecting with the fundamental theorems of logic, this paper will use the language of category theory to explore objective and subjective problems, problems of information and receivers, problems of complexity and interpreters. In this paper, an abstract functor as an interpreter is introduced as a crucial tool to observe a certain object from a formal system. A certain proposition being provable or disprovable is corresponding to whether it is PAC learnable or not from the perspective of an interpreter, thus a more general framework which contains the Gödel incompleteness theorem could be built.

This study provides the general Gödel incompleteness theorem (GIC) and gives its dimension calculation. These results are then used to (a) demonstrate Cantor diagonal proof, original Gödel incompleteness theorem, Turing halting problem under the same framework; (b) give the critical state a mathematical description as a Gödel incompleteness space which can be applied to a variety of studies; (c) explain the spin glass state as a “frozen” incomplete state to explain why glass is transparent, and provide possible calculation formulas for the energy band; (d) corresponds the incompleteness dimensions to the dimensions in fundamental physics.

## *1 Complexity Measurement Based on Interpreters*

First, we construct the process of complexity measurement using the language of category theory. Next, we generalize the PAC learnable theory for the interpreters. Lastly, we find the Gödel incompleteness theorem for this generalized PAC learnable theory.

Definition : System

System $S$ is defined as a collection containing objects $A$, morphisms between objects $L$, and environments $(E)$. $S:=S\,(A,L,E)$. The environment identifies the elements of the system. Elements constitute objects. Objects and the morphisms between objects form the structure of the system. Let the environment E be an object of a locally small category, e.g., a set, then the system

S is locally small. $A$ is the collection of all well-constructed strings, which could be considered as the free monoid of environment $E$ module some construction rule $R$ that if one string is not well-constructed, it is seen as the empty string with length 0, i.e., the initial object of $S$. Morphisms $L$ has some certain rules to map from one object to another which is not necessarily to be the monoid homomorphism. $E$ could be described as an automorphism of category $S$.

Definition : Information

Information $\mathcal{A}$ is defined as an object being mapped from another object by an interpreter. Interpreter $\varphi$ is defined as a functor from one system S to another system C satisfying $S \overset{\varphi}{\rightarrow} C$ and $\varphi(A) = \mathcal{A}$.

Definition : the amount of information

The amount of information $I$ is determined by both the information itself and the interpreter. The traditional information amount is $I(A) = \log|\Phi(A)|$ where $\Phi(A)$ is the set of all possible strings with the same length of $A$. $|\Phi(A)|$ means counting the number of elements of $\Phi(A)$. With the interpreter the information amount should be $I(\varphi(A)) = \log\left(\frac{|\Phi(A)|}{|\varphi(A)|}\right)$, where $\varphi(A)$ means the set of all the possible strings that $A$ is interpreted as with a possibility distribution and $|\varphi(A)|$ means counting the number of elements of $\varphi(A)$. If $A$ only has one possible interpretation, $I(\varphi(A)) = \log|\Phi(A)|$ and this will be the biggest information amount. If $A$ is interpreted to be all possible strings, $I(\varphi(A)) = \log 1 = 0$ and this will be the smallest information amount, which means one gets nothing from $\varphi(A)$.

More generally, let $\boldsymbol{H}$ represent the set (category, system) of all interpreters (hypotheses, heuristic hypotheses). $\boldsymbol{h}$ is a specific interpreter (morphism, mapping, operator) in $\boldsymbol{H}$. Let all the information amount calculation be integrated into $h$. The generalized information amount I is defined as follows.

$$I(a \leq x \leq b) = \int_a^b h(A)(x)dx$$

Where $A$ is the original object. $h$ is the interpreter. $h\ (A)$ is the distribution, which means it can be normalized, as a whole. $s$ is the independent variable of the distribution from a cardinal category Λ. $a$ and $b$ are the upper and lower bounds of $x$. Λ is not necessarily a partial-ordered set, but we can give it an order to get the part of information we need without loss of generality.

Definition : complexity

The complexity $c$ of the information $h(A)$ is defined as follows:

$$\textbf{continuous form}: \ c(h) = \int_a^b s(x) \ \cdot [h(A)]\big(s(x)\big)dx$$

$$\textbf{dicscrete form}: \ c(h) = \sum_{i=a}^{b} s_i \ \cdot h(A)(s_i)$$

Let $h(A)$ be a distribution, which means it could be normalized. Then $c(h)$ could be considered as a kind of expectation and $s$ will be measurement or possibility. Then $c(h)$ will satisfy all the property of expectation. For example, taking $h(A)$ as $-log_2$ and $s$ as $p_i$, (or $(i, p_i)$ if we need to distinguish same $p_i = p_j$ and $i \in \Lambda$) we can get the Shannon information entropy, which indicates the expectation of information increase by adding a new bit, i.e., the average information amount per bit. Our generalized complexity measurement indicates the expectation of information increase by adding a new position of cardinal, i.e., the average information amount per cardinal, which is in line with the Shannon entropy.

## *2 PAC learnable interpreter*

Definition 5: PAC learnable for interpreters

We set up the learning conditions as follows:

$$f \ := the\ function\ we\ want\ to\ learn, the\ target\ function$$

$$F \coloneqq the\ class\ of\ functions\ from\ which\ f\ can\ be\ selected. f \in F$$

$$I \coloneqq the\ set\ of\ all\ possible\ individuals.$$

$$X \coloneqq X\ is\ the\ domain\ of\ f$$

$$N \coloneqq the\ cardinality\ of\ X.\ \ i.e. N = |X|$$

$$D := a\ (probability)\ disturbution\ on\ X\ used\ for\ training\ and\ testing$$

$$H \ := \ the\ set\ of\ possible\ interpreters\ (hypotheses)$$

$$h\ is\ the\ specific\ interpreter\ (hypothesis)\ that\ has\ been\ learned. h \in H$$

$$m \coloneqq the\ cardinality\ of\ the\ training\ set$$

Where h is considered as an operator.

Let $EX(c, D)$ be a procedure that draws an example $x$, using a probability distribution $D$ and gives the correct label $c(x)$, that is 1 if $x \in c$ and 0 otherwise.

E.g., Let $h(I) = D = \lambda||\psi\,||^2 = ||\hat{\psi}\,||^2$ where $\sum||\hat{\psi}\,||^2 = 1$ and $\sum||\psi\,||^2 = \frac{1}{\lambda} < +\infty$. On this condition, we can consider $D$ as a possibility distribution.

Note that we can examine learning in terms of functions or of concepts, i.e. sets i.e. interpreters. They are equivalent, if we remember the use of characteristic functions for sets.

Now, given $0 < \epsilon < 1, 0 < \delta < 1$, assume there is a cognitive algorithm (a decomposition algorithm) (an algorithm $A$) and a polynomial $p$ in $\frac{1}{\epsilon}, \frac{1}{\delta}$ (and other relevant parameters of the class $C$) such that, given a sample of size $p$ drawn according to EX(c, D), then, with probability of at least $1 - \delta$, $A$ outputs a interpreter $h \in C$ that has an average error less than or equal to $\epsilon$ on $I$ with the same distribution $D$. Further if the above statement for algorithm $A$ is true for every concept $c \in C$ and for every distribution $D$ over I and for all $0 < \epsilon, \delta < 1$, then $C$ is (efficiently) **PAC learnable** (or distribution-free PAC learnable). We can also say that $A$ is a **PAC learning algorithm** for $C$.

Write the discussions above in formula:

$A$ is a **PAC learning algorithm** for $C$ if:

$$P1: \exists\ wave\ function\ \psi\ s.t.$$

$$h(I) = \lambda||\psi\,||^2 = ||\hat{\psi}\,||^2 \Leftrightarrow h(I) = D\ is\ a\ possibility\ distribution$$

$$and$$

$$P2: n\ is\ greater\ than\ a\ certain\ number\ N$$

$$n > N = \frac{ln\frac{2|H|}{\delta}}{2\epsilon^2}$$

$$\Leftrightarrow$$

$$P3: \forall c \in C, \forall D\ over\ I\ and\ \forall 0 < \epsilon, \delta < 1, \exists h\ output\ by\ A\ s.t.$$

$$I\left\{\left|\widehat{c(h)} - c(h)\right| \leq \epsilon\right\} \geq 1 - \delta$$

$$\Leftrightarrow$$

$$P4: A\ is\ PAC\ learnable.$$

where the definitions:

1.

Information (of I under the interpreter h) or we can call it partial information

$$I_h\{a \le f \le b\} \equiv I\{a \le f \le b\} = \int_a^b h(I)(s)ds$$

Where $a$ and $b$ in cardinal category $\Lambda$ are the upper and lower bounds of $x$.

$a$ and $b$ could be any lower bound and upper bound which is allowed to be integrated on.

2.

The amount of information

$$\left||I|\right| = I_h\{f_{min} \le f \le f_{max}\} \equiv I\{f_{min} \le f \le f_{max}\} = \int_{f_{min}}^{f_{max}} h(I)(s)ds$$

3.

Complexity (of I under the interpreter h) or we can call it First-order origin (moment) complexity

$$c = c(I_h) \equiv c(h) = \int_a^b s \cdot h(I)(s)ds$$

4.

N-order origin complexity

$$c^N = c^N(I_h) \equiv c^N(h) = \int_a^b s^N \cdot h(I)(s)ds$$

N-order central complexity

$$\mathcal{D}^N = \mathcal{D}^N(I_h) \equiv \mathcal{D}^N(h) = \int_a^b (s-c)^N \cdot h(I)(s)ds$$

5. $\left|\widehat{c(h)} - c(h)\right|$ is the loss value of the complexity measurement where $\widehat{c(h)}$ is the predicate value given by algorithm A and $c(h)$ is the real value inherited in interpreter $h$ itself.

Here we prove the boundary $n > N = \frac{ln\frac{2|H|}{\delta}}{2\epsilon^2}$ by proving the following lemmas.

**(1) The Markov inequality proved under the framework of this text**

Let f⩾0 be the independent variable of a non-negative interpreted distribution h(A), with the complexity of $h(A)$ being $c$. For any $t > 0$, we have:

$$I(f \geq t) \leq \frac{c}{t}$$

Pf：

$$I(f \geq t) = c\left[1_{\{f\geq t\}}\right] \leq c\left[\frac{f}{t}1_{\{f\geq t\}}\right] \leq \frac{c}{t}$$

Where $1_{\{f\geq t\}}$ represents if $f \geq t$ map $1_{\{f\geq t\}}$ to 1and 0 otherwise. So, under distribution h(A), there is $1_{\{f\geq t\}} \leq 1$.

■

**(2) The Chebyshev's inequality proved under the framework of this text**

For any $\epsilon, I$ is the amount of information, $\mathcal{D}^2$is the 2-order central complexity, $c$ is the first-order original complexity, we have

$$I(|f-c| \geq \epsilon) \leq \frac{\mathcal{D}^2}{\epsilon^2}$$

Pf：

$$I(|f-c| \geq \epsilon) = \int_{f\leq c-\epsilon}^{f\geq c+\epsilon} h(A)(f)df$$
$$\leq \int_{f\leq c-\epsilon}^{f\geq c+\epsilon} \left(\frac{|f-c|}{\epsilon}\right)^2 h(A)(f)df$$
$$\leq \int_{-\infty}^{+\infty} \left(\frac{|f-c|}{\epsilon}\right)^2 h(A)(f)df$$
$$= \frac{\mathcal{D}^2}{\epsilon^2}$$

■

**(3) The Chernoff's bound proved under the framework of this text**

Let $f \in \mathbb{R}$ be the independent variable of the interpreted distribution $h(A), c$ is the first-order original complexity. For any $t > 0$， $s > 0$, we have：

$$I(f \geq t) \leq \frac{c(e^{sf})}{e^{st}}$$

Pf：

$$I(f \geq t) = I(sf \geq st) = I(e^{sf} \geq e^{st})$$

$$\leq \frac{c(e^{sf})}{e^{st}}$$

■

**(4) The Hoeffding's lemma proved under the framework of this text**

Let $Z \in [a, b]$ be the independent variable of the interpreted distribution $h(A), c$ is the first-order original complexity. For any $\lambda \in \mathbb{R}$, we have：

$$c\left[\exp\left(\lambda\big(Z - c(Z)\big)\right)\right] \leq \exp\left(\frac{\lambda^2(b-a)^2}{8}\right) \qquad (2-18)$$

Pf：Let $c(Z) = 0$ to make the proof more neat and without loss of generality. Setting $c(Z)$ be any number is fine.

Only need to prove,

$$c\big[\exp\big(\lambda(Z)\big)\big] \leq \exp\left(\frac{\lambda^2(\mathrm{b}-\mathrm{a})^2}{8}\right)$$

Let $Z = \alpha b + (1-\alpha)a$, where $\alpha = \frac{Z-a}{b-a}$, $1-\alpha = \frac{b-Z}{b-a}$. Let $g(Z) = \exp(\lambda Z)$, since $g''(Z) > 0$, so we can get：

$$g(Z) = g(\alpha b + (1-\alpha)a) \leq \alpha g(b) + (1-\alpha)g(a)$$

That is,

$$g(Z) \leq \frac{Z-a}{b-a}\exp(\lambda b) + \frac{b-Z}{b-a}\exp(\lambda a)$$

Taking the complexity on both sides of the inequality gives:

$$c[\exp(\lambda Z)] \le c[\frac{Z-a}{b-a}\exp(\lambda b) + \frac{b-Z}{b-a}\exp(\lambda a)]$$

$$= c\left[\frac{Z}{b-a}(\exp(\lambda b) - \exp(\lambda a))\right] + c[\frac{b}{b-a}\exp(\lambda a) - \frac{a}{b-a}\exp(\lambda b)]$$

And since $c(Z)$=0, therefore,

$$c[\exp(\lambda Z)] \le c[\frac{b}{b-a}\exp(\lambda a) - \frac{a}{b-a}\exp(\lambda b)]$$

Let $\gamma = -\frac{a}{b-a}$. We have $1-\gamma = \frac{b}{b-a}$,and $a$ and $b$ are constants, then the inequality can be simplified to

$$c[\exp(\lambda Z)] \le c[(1-\gamma)\exp(\lambda a) - \gamma\exp(\lambda b)] = (1-\gamma)\exp(\lambda a) - \gamma\exp(\lambda b)$$

$Let\ \mu = \lambda(b-a)$, we have $\lambda a = -\mu\gamma$, then the inequality can be simplified to

$$c[\exp(\lambda Z)] \le \exp(-\mu\gamma)(1-\gamma+\gamma\exp(\mu))$$

Let $f(\mu) = \log[\exp(-\mu\gamma)(1-\gamma+\gamma\exp(\mu))]$, then we have $c[\exp(\lambda Z)] \le \exp[f(\mu)]$Derived from $f(\mu)$：

$$f'(\mu) = -\gamma + \frac{\gamma\exp(\mu)}{1-\gamma+\exp(\mu)}$$

$$f''(\mu) = \frac{\gamma(1-\gamma)\exp(\mu)}{1-\gamma+\exp(\mu)}$$

According to Taylor's theorem, there exist a $\xi \in (0,\mu)$， so that： $f(\mu) = f(0) + \mu f'(0) + \frac{\mu^2}{2}f''(\xi)$ is satisfied. It is known that $f(0) = 0, f'(0) = 0$, therefore

$$f(\mu) = \frac{\mu^2}{2}f''(\xi)$$

Let $t = \frac{\gamma\exp(\mu)}{1-\gamma+\exp(\mu)}$, so we have $f''(\xi) = t(1-t) \le \frac{1}{4}$, therefore

$$f(\mu) \le \frac{\mu^2}{8} = \frac{\lambda^2(b-a)^2}{8}$$

From the definition of $f(\mu)$ we could get

$$c\left[\exp\left(\lambda(Z)\right)\right] \le \exp\left(\frac{\lambda^2(\mathrm{b}-\mathrm{a})^2}{8}\right)$$

■

**(5) The Hoeffding inequality proved under the framework of this text**

Let $I$ be the information amount. Let there be $N$ independent variables $Z_i \in [a,b], -\infty < a \le b < \infty$, that are independently distributed after interpretation. For $t > 0$, we have

$$I\left(\frac{1}{N}\sum_{i=1}^{N}(Z_i - c(Z_i)) \ge t\right) \le \exp\left(-\frac{2Nt^2}{(b-a)^2}\right) \qquad (2-19)$$

Pf：

$$I\left(\frac{1}{N}\sum_{i=1}^{N}\left(Z_i - c(Z_i)\right) \ge t\right) = I\left(\sum_{i=1}^{N}\left(Z_i - c(Z_i)\right) \ge Nt\right)$$

From **Chernoff's bound** let $s > 0$ so that

$$I\left(\sum_{i=1}^{N}\left(Z_i - c(Z_i)\right) \ge Nt\right) \le \frac{c\left[\exp\left(s\sum_{i=1}^{N}\left(Z_i - c(Z_i)\right)\right)\right]}{\exp(sNt)}$$

$$= \frac{\prod_{i=1}^{N} c\left[\exp\left(s\left(Z_i - c(Z_i)\right)\right)\right]}{\exp(sNt)}$$

Simplify this by **Hoeffding's lemma**：

$$I\left(\frac{1}{N}\sum_{i=1}^{N}\left(Z_i - c(Z_i)\right) \ge t\right) \le \frac{\prod_{i=1}^{N} c\left[\exp\left(s\left(Z_i - c(Z_i)\right)\right)\right]}{\exp(sNt)}$$

$$\le \frac{\prod_{i=1}^{N} \exp\left(\frac{s^2(b-a)^2}{8}\right)}{\exp(sNt)}$$

$$= \exp\left[\frac{Ns^2(b-a)^2}{8} - sNt\right]$$

$Let\ h(s) = \frac{Ns^2(b-a)^2}{8} - sNt$ be the quadratic function of $s$ and $s > 0$. It can be obtained by finding the minimum value based on the properties of quadratic functions

$$\min h(s) = -\frac{2Nt^2}{(b-a)^2}$$

Therefore, to ensure that the inequality holds for any $s$, taking the minimum of $h(s)$ to reach the upper bound at this time, we can obtain:

$$I\left(\frac{1}{N}\sum_{i=1}^{N}\big(Z_i - c(Z_i)\big) \geq t\right) \leq \exp\left(-\frac{2Nt^2}{(b-a)^2}\right)$$

■

The **Hoeffding inequality** gives out the bound of the size of the sample set。

$$\begin{aligned} P[\exists h \in \mathcal{H}, |\hat{c}_S(h) - c(h)| \geq \epsilon] &= P\left[\bigvee_{h_i \in \mathcal{H}} |\hat{c}_S(h_i) - c(h_i)| \geq \epsilon\right] \\ &\leq \sum_{h_i \in \mathcal{H}} P[|\hat{c}_S(h_i) - c(h_i)| \geq \epsilon] \\ &= 2|H| \exp(-2\epsilon^2 N) = \delta \end{aligned}$$

Therefore, for the size of the training set $n$, such $n > N = \frac{ln\frac{2|H|}{\delta}}{2\epsilon^2}$ satisfies the PAC learnable condition.

Proof of this theorem is trivially a generalization of the original PAC learnable condition.

Definition: general PAC (GPAC) learnable

Define that A is general PAC (GPAC) learnable algorithm iff:

$$P1': \exists\ wave\ function\ \psi\ s.t.$$

$$h(I) = \lambda||\psi||^2 = \lambda'||\hat{\psi}||^2 \Leftrightarrow h(I) = \frac{1}{\lambda'}D\ is\ a\ possibility\ distribution$$

$$where\ \ \lambda' = \big||I|\big| = \int_{f_{min}}^{f_{max}} h(I)(s)ds$$

$$and$$

$$P2': n\ is\ greater\ than\ a\ certain\ number\ N$$

$$n > N = \frac{ln\frac{2|H|}{\delta}}{2\epsilon^2}$$

$$\Leftrightarrow$$

$$and\ P3': \forall c \in C, \forall D\ over\ X, and\ \ \forall 0 < \epsilon\ , \delta\ < 1\ , \exists h\ output\ by\ A\ s.t.$$

$$I\{\left|\widehat{c(h)} - c(h)\right| \le \epsilon\} \ge \lambda'(1-\delta)$$

$$\Leftrightarrow$$

$$P4': A\ is\ general\ PAC\ learnable.$$

In the next parts, we could mix PAC and general PAC, which is smoothly acceptable. When the total amount of information could be normalized, the mixing of the two does not affect whether or not it's learnable. In fact:

$$P1' \Leftrightarrow -\infty < \|I\| < +\infty$$

As long as $\lambda' \neq 0$ exists, P1 is satisfied.

## *3 Completeness of interpreter for proposition*

For

X is an object from an outside system (e.g. real world).

$\varphi$ is an interpreter.

$I$ is the information one cognitive system gets.

H is an operator which follows the axioms of probability.

S is the decomposition of our comprehension (information on each eigenvector is exposed).

S is on the other hand the probability of a particle.

$\psi$ is a wave function.

$$(\varphi:)X \longrightarrow (H:)I \rightarrow S \leftarrow \psi(:norm)$$

This equation is an illustration of the process of cognition. We emphasize that the process of an external object *X* being recognized will first be processed by the interpreter $\varphi$ into information *I* that we can process, and then *I* will get a result *S* that carries meaning after processing by the interpreter $H$. This meaning $S$ is characterized by a distribution and thus corresponds to a wave function $\psi$, which differs from the probability distribution by a normalized norm. We assume that the steps and results of our cognitive process $S$ can be described by a formal system. The formal system means that our cognitive process is a logical process, and the formal system is exactly the formal description of the logical process.

If H is general PAC learnable, we can get:

$$HI = S = \lambda|\psi|^2$$

$$\hat{H}|I\rangle = |S\rangle = \lambda \sum_i \langle e_i|\psi\rangle \langle\psi|e_i\rangle|e_i\rangle$$

Where:

H is the space of interpreters.

I is the information (after we observe X by $\varphi$).

S is a formal system, containing all the information after we decompose $I$ by $H$.

For example, X is a particle. I is the position information of X. $H_1$ is sampling in the time domain, in other words, delta function sampling, then $S_1$ is the expansion in the time domain. $H_2$ is sampling in the frequency domain, in other words, Fourier transformation sampling, then $S_2$ is the expansion in the frequency domain. Other understandings can also include the coordinate momentum representation and the particle number representation in quantum mechanics. Decomposition of $I$ by $H$ means we choose a certain representation. Let $\psi$ be a wave function.

$$|\psi\rangle = \begin{pmatrix} c_1 e_1 \\ c_2 e_2 \\ \cdots \end{pmatrix},\ c_i \in \mathbb{C},\ \{e_i\}\ is\ a\ set\ of\ normal\ orthognal\ eigen\ basis.$$

$$\langle 1|\psi\rangle = \Sigma \mathrm{c_i e_i} = \int dk\, c_k e_k$$

### 3.1 Definition of Interpreter Space H

For $A_i \in H$, $A_i$ is often considered as an operator (a matrix) and $A$ is constructed as a subspace of $H$. We define $A = \{A_i|\ \ A_i\ is\ PAC\ learnable\}$

Lemma 1: the linear space for interpreters

1.Addition

$$A_i, A_j \in A \Rightarrow A_i + A_j \in H$$

Proof:

$$(A_i + A_j)|I\rangle = s_{A_i} + s_{A_j} \sim \lambda \left\| \psi_{A_i} + \psi_{A_j} \right\|^2$$

$s_{A_i} + s_{A_j}$ could be considered as the distribution of the sum of two random variable.

So $A_i + A_j$ is PAC learnable.

So $A_i + A_j \in H$.

■

2.Scaling

$$A_i \in H \Rightarrow \lambda_i A_i \in H$$

Pf:

$$\lambda_i A_i|I\rangle = \lambda_i s_{A_i} \sim \lambda_i \lambda_{A_i} \left\| \psi_{A_i} \right\|^2$$

$\lambda_i A_i$ could be considered as $\lambda_i$ of $\psi_{A_i}$ wave comes together.

So $\lambda_i A_i$ is general PAC learnable.

So $\lambda_i A_i \in H$.

■

3.Operator on operator is allowed.

Let B be an operation working on $A_i$, i.e. $BA_i \in H$. We only allow the $B$ that guarantee the **closure** of $BA_i$.

So, the entire space H could be described as an expansion from a basis $H_0$. A basis $H_0$ in $H$ is corresponding to the axioms in $S$ with regard to a fixed $I$. Based on the basis of H, $H_0$ is inherited in H that belongs to H. Then we can expand $H_0$ to reconstruct H as the following:

$$H = span(BH_0) = span(B)H_0$$

The operation of this step is equivalent to allowing higher order operators, which is corresponding to allowing high order predicate operation.

4.Tensor Space

If $A_i, A_j \subseteq H$ and $A_i \wedge A_j = \emptyset$, we define $A_i \otimes A_j \in H_{\otimes}$ is the Tensor Space of H.

$$A_i \otimes A_j |I\rangle = s_{A_i} \otimes s_{A_j} \in S$$

The definition of this tensor product corresponds to quantum entanglement, which allows us to discuss about the entanglement phenomenon of cognitive process. For example, an entanglement of information A and B means that knowing A means knowing B regardless of time or space distance and vice versa. In logic, this is corresponding to the equivalent proposition. The observation process of entangled particles is just corresponding to the entanglement phenomenon of information process.

### 3.2 Definition of Formal system S

Definition 7：Formal system S

Let S be a system which contains all the objects mapped by $AI$.

One system has three components containing Objects, Morphisms between objects and Environment.

System S=S($\mathcal{A}$,L,E) which is a triplet. $\mathcal{A}$ is the set of objects. L is the set of morphisms (links). E is the environment indicating the elements that could compose objects.

Objects:= well-constructed string composed by elements in Environment

For $H|I\rangle = S$

Well-constructed rules are implied by space H. Specifically, if one certain operation is allowed, this operation indicating well-constructed way of combination guides how the elementary elements compose an object. Compared with formal systems in traditional logic, objects are propositions, morphisms are logic, and environments are elementary sentence.

Morphism: = relationship between Objects. Examples:

$L(A1) = \{A2, A3\}$

$\mathrm{Map}(\mathrm{A1}, \mathrm{A1} \rightarrow \mathrm{A2}) = \mathrm{A2}$

Environment: = elements that could compose Objects.

Everything that could be decomposed from H and I is in the Environment.

Original formal system only contains objects (originally called propositions) and morphisms (originally called logic procedure). We introduce the concept of environment so that we can involve the basic elements that make up an object (a well-formed string) in the discussion of the amount of information and complexity of that object, so we have to include the concept of environment when defining the system. For example, when we talk about DNA systems, we're talking about DNA sequences, and the relationships between DNA and DNA are defined as morphisms, like how to get from one sequence to another by splicing through some operation and so on. When we consider the codon (three-digit DNA) as the minimum discussion scale of DNA and when we consider the one-digit DNA as the minimum discussion scale of DNA, we calculate the amount of information and complexity of the same DNA sequence completely different. Here the two different environments correspond to E1: 61 effective codons and E2: 4 deoxyribonucleic acid AGCT.

For example, the string GCAGCG, the E1 environment is equivalent to the object structure GCA-GCG, the calculated information is $\log_2 61^2$. The E2 environment is equivalent to the object structure G-C-A-G-C-G, and the calculated information is $\log_2 4^6 = 12$.

So even if we're talking about the exact same relationship between objects, in other words, the same thing from a categorical point of view, we still get different amounts of information because of the way we look at it, so it makes sense to attach to a category a set of its smallest constituent elements and this definition of the law of well-constructed combination, and we call it a system. The definition of a system is equivalent to a category $H$ with an outer category $I$ which has a functor $I \to H$, the object of the outer category indicating the basic elements that make up the object of the inner category, and the functor indicating a well-constructed (operation of combining elements legally). The “environment” that we define here is closer to “context” than the external environment.

For defining system $S = H|I\rangle$, S is equivalent to a category $H$ with a outer category $|I\rangle$ and functor $|I\rangle$ acting on $H$.

Lemma : We can construct the linear space for the formal system.

By definition:

1.Addition

$$s_i, s_j \in S \Rightarrow s_i + s_j \in S$$

Pf:

Element + and number $\lambda$ are components of space H.

$$\{+\} \in S(E)$$

Proof in diagram is shown in Fig. S1. And linear operations “plus two vectors” and “times number $\lambda$” are within the operations of H.

So $s_i + s_j$ is well-defined.

So $s_i + s_j \in S$

Specifically in category theory, we define addition as following, so that we can make $s_{h_1+h_2}$ and $s_{h_1} + s_{h_2}$ equivalent by constructing the morphism "def".

■

2. Scaling

Let $\lambda h_1 \sim \sum_1^{\lambda} h_1$, where ~ is equivalent.

Then we can define scaling as in the Fig. S2.

We can make $s_{\lambda h_1}$ and $\lambda s_{h_1}$ equivalent by constructing the morphism "def".

3.Operator on operator is allowed.

Let $B$ be an operation working on $A \in H$, i.e. $BA_i \in H$

Let $\tilde{B}$ be the operator on $S$ so that make the diagram in Fig. S3 commute.

Then we don't need to difference the $B$ and $\tilde{B}$.

Then $BA|I\rangle = B(s_A) = s_{BA}$. In other words, we define operator on operator in Fig. S4.

Pf:

Operator B is one component of space H and B indicates well-constructed string.

So $BA|I\rangle = B(s_A) = s_{BA}$ is a well-constructed string in S(A,L,E).

So $s_{BA} \in S$

So the entire space S could be described as an expansion from a basis $S_0$. A basis $H_0$ in $H$ is corresponding to the axioms in $S$ with regard to a fixed $I$. Based on the basis of S, $S_0$ is inherited in S that belongs to S. Then we can expand $S_0$ to reconstruct S as the following:

$$H = span(BH_0) = span(B)H_0$$

$$H|I\rangle = span(B)S_0$$

4.Tensor Space

If $A_i, A_j \subseteq H$ and $A_i \wedge A_j = \emptyset$, we define $A_i \otimes A_j \in H_{\otimes}$ is the Tensor Space of H.

We have:

$$A_i \otimes A_j|I\rangle = S_{A_i} \otimes S_{A_j} \subseteq S_{\otimes}$$

Pf:

$\otimes$ this operation is allowed in H s.t, two subspace of H could be mapped to another set. So $\otimes$ is an element of S(E) and this link between $s_{A_i}$ *and* $s_{A_j}$ is well-defined.

So $s_{A_i} \otimes s_{A_j} \in S$

So $S_{A_i} \otimes S_{A_j} \in S_{\otimes}$

■

The category diagram of the definition of the tensor space is in Fig. S5.

### 3.3 True propositions and False propositions

In this section, we define the corresponding truth-falsity for the PAC learnable interpreter system.

As the Equivalent Equation (3.1) said, if H is a GPAC learnable space, there exists a correspondence between interpreter space H, formal system S and quantum system Φ.

The H is corresponding to the proof process.

The S is containing each proposition.

So Φ is defined to determine the "True or False".

Define:

Proposition $s \in S$ is true iff $s = \lambda|\psi|^2$ has $\lambda > 0$

Proposition $s \in S$ is false iff $s = \lambda|\psi|^2$ has $\lambda < 0$

If $s = \lambda|\psi|^2$ has $\lambda = 0$, s is an empty proposition. And if $s = \lambda|\psi|^2$ has $\lambda = \infty$, then H is not GPAC learnable, and $s \notin S$.So, this definition about "True or False" is well-defined.

We call this definition the normal "True or False". In fact, this is just one of the possible definitions of the "True or False". For example, $Re(\lambda) \geq 0, Re(\lambda) \leq 0 , \lambda = 0, \lambda \to \infty$ could also be a well-defined definition of the "True or False". If we do not specially say, the normal "True or False" is used. One can see:

$$s \in S\ and\ if\ s\ is\ true \Rightarrow \neg s = -s \in S\ and\ \neg s\ is\ false$$

### 3.4 The Space of Logic

Interpreter $h \in H$ is an operator working over the information $I$. And if there is an operator $B \in \{H \rightarrow H\} \sim H^2$ , we get $Bh \in H$ which is another interpreter. This is what we've discussed in the definition of H.

When we consider $H$ is the space of all logical procedure, in other words, a proof process, we can change the meanings of $H, I$ and $H^2$ corresponding to a logical system.

Let:

I be the axioms of a calculation process, e.g., **Peano Arithmetic**.

$H$ be the space of all legal logical procedure

$B$ be all legal procedure on H forming space $H^2$

$BH$ can be characterized by $BA_0$ where $A_0$ is the axioms of $H^2$

[i.e., $A_0$ are the eigen vectors of the space $H^2$.]

And $H^2$ could also have enough eigen vector $B_0\ s.t, span(B_0) = H^2$

So $H = span(A_0), H^2 = span(B_0)$

So, we have

$$S = HI = span(A_0)I = S_{span(A_0)} = span(B_0)A_0I$$

$S_{A_0}$ also works like the eigen vectors of S.

But as we can see next,

$$H_{\otimes}I = (H \otimes H)I = span(B_0)[span(A_0) \otimes span(A_0)]I = S \otimes S \neq S_{\otimes}$$

By definition: For $B_1, B_2 \in \{H \rightarrow H\} \sim H^2, A_1, A_2 \in H$

$$B_1A_1 + B_2A_2 \in H$$

We have

$$(B_1A_1 + B_2A_2)I = B_1s_{A_1} + B_2s_{A_1} = s_{B_1A_1} + s_{B_2A_2} \in S$$

Therefore, **there is a bijection so that $\boldsymbol{H} \leftrightarrow \boldsymbol{S}$**. The bijection $\boldsymbol{H} \leftrightarrow \boldsymbol{S}$ indicates the original Gödel completeness theorem when considering H as proof procedure and S as propositional formal system. The diagram shows as in Fig. S6.

$$\forall s_X \in S, \exists X \in H\ s.t. XI = s_X \Leftrightarrow S\ is\ complete$$

$$\exists s_X \in S, \nexists X \in H\ s.t. XI = s_X \Leftrightarrow S\ is\ incomplete$$

For $B_1, B_2 \subseteq \{H \to H\} \sim H^2, A_1, A_2 \subseteq H$, we define $\neg A = H - A, \neg B = \{H \to H\} - B$. We have:

$$(B_1 \otimes (\neg A_1) + (\neg B_2) \otimes A_2)I = B_1 S_{\neg A_1} + \neg B_2 S_{A_1} = S_{B_1 \otimes (\neg A_1)} + S_{(\neg B_2) \otimes A_2} \in S_{\otimes}$$

But

$$B_1 \otimes (\neg A_1) + (\neg B_2) \otimes A_2 \in H_{\otimes}$$

$$\Leftrightarrow$$

$$\exists H_1, H_2 \in H\ s.t. H_1 \otimes H_2 = B_1 \otimes (\neg A_1) + (\neg B_2) \otimes A_2$$

Most of the time, entangled polynomial of elements cannot be decomposed by two $H_1, H_2 \in H$, in other words, $B_1 \otimes (\neg A_1) + (\neg B_2) \otimes A_2 \in H_{\otimes}$ isn't always correct.

But $S_{B_1 \otimes (\neg A_1)} + S_{(\neg B_2) \otimes A_2} \in S_{\otimes}$ is always correct.

**There may not be a bijection so that $\boldsymbol{H}_{\otimes} \leftrightarrow \boldsymbol{S}_{\otimes}$**

## *4 Incompleteness of interpreter space for proposition space*

In this section, we construct the Gödel Proposition Space firstly, which indicates that a certain sentence in this space may cause the failure of the bijection. And then we find this certain sentence using diagonal proof to finish the whole proof of the incompleteness of interpreter space for proposition space, which gives the Gödel incompleteness theorem for PAC learnable theory.

### 4.1 Gödel Proposition Space $\boldsymbol{S}_{\boldsymbol{G}\ddot{o}\boldsymbol{del}}$

Here we are going to prove that there exists some object $\boldsymbol{S}_{\boldsymbol{G}\ddot{o}\boldsymbol{del}}$ in $S_{\otimes}$ that does not have the corresponding object in $H_{\otimes}$ so that $HI = \boldsymbol{S}_{\boldsymbol{G}\ddot{o}\boldsymbol{del}}$ . This conclusion is equivalent to the incompleteness theorem under this framework. Now we're going to prove this **by contradiction.**

Inspired by quantum entanglement, let's consider a candidate object in the tensor system $S_{\otimes}$:

$$\boldsymbol{S_{Gödel}} = S_B \otimes S_A + S_{\neg B} \otimes S_{\neg A}$$

Where $S_B \ and \ S_A$ are two sub-systems of S, which means the corresponding interpreter space $A \subseteq H$ and A is subspace of H while $B \subseteq H^2$ and B is a subspace of $\{H \to H\}$. We specially claim:

$$span(B \otimes A) = H_{\otimes}$$

As $\forall \ h \in H, \exists b \in B \ and \ a \in A \ s.t, b(a) = h$

$\boldsymbol{S_{Gödel}}$ is well-defined.

Let's find the corresponding tensor in $H_{\otimes}$ . If one exists, it should look like this:

$$H_G = B \otimes A + (\neg B) \otimes (\neg A)$$

Here we give the proof that $\boldsymbol{S_{Gödel}}$ **doesn't have the corresponding** $\boldsymbol{H_G} \in \boldsymbol{H_{\otimes}}$ by contradiction:

Hypothesis "each element in $S_{Gödel}$ could be written as $H_{G_{ij}}I$"

Suppose

$$H_G \in H_{\otimes} \iff \exists H_1, H_2 \ s.t, H_1 \otimes H_2 = B \otimes A + (\neg B) \otimes (\neg A)$$

Constructed general form, when $\mathbb{F}$ is a field in mathematics：

$$H_N = aB \otimes A + bB \otimes (\neg A) + c(\neg B) \otimes A + d(\neg B) \otimes (\neg A), a, b, c, d \in \mathbb{F}$$

If $H_N \in H_{\otimes}$, which needs to satisfy determination $\Delta$:

$$\Delta = \begin{vmatrix} a & b \\ c & d \end{vmatrix} = ad - bc = 0$$

When $B$ and $\neg B, A$ and $\neg A$ are all linear independent, in other words, $B \otimes A$ and $(\neg B) \otimes (\neg A)$ cannot be combined with items of the same type, for $H_G$， we can get $\Delta = 1$, hence $H_G \notin H_{\otimes}$. So suppose each object in $S_{Gödel}$ can be written in this form of $H_{G_{ij}}I$, we have:

$$\begin{aligned} & dimB \neq dimH^2 \ or \ dimA \neq dimH \\ & s.t, H_G \notin H_{\otimes} \end{aligned}$$

This is a contradiction, $H_G \in H_{\otimes} \rightarrow\leftarrow H_G \notin H_{\otimes}$. So our hypothesis "each element in $S_{Gödel}$ could be written as $H_{G_{ij}}I$" is incorrect. Hence "there exists some object $\boldsymbol{S_{Gödel}}$ in $S_{\otimes}$ that does not have the corresponding object in $H_{\otimes}$ so that $HI = \boldsymbol{S_{Gödel}}$" is correct. In other words, a bijection $H_{\otimes} \leftrightarrow S_{\otimes}$ doesn't always exist. If This proves the incompleteness between $H_{\otimes}$ and $S_{\otimes}$.

In conclusion, the theorem "**Incompleteness of interpreter space for proposition space" goes as following:** if H satisfies the following properties:

$$\exists B\ is\ the\ subspace\ of \{H \to H\},$$
$$\exists A\ is\ the\ subspace\ of\ H$$
$$s.t, span(B)span(A) = H$$

$$and\ dimB < dimH^2\ or\ dimA < dimH$$

Then such bijection $H_{\otimes} \leftrightarrow S_{\otimes}$ doesn't exist. This relationship is shown in Fig. S7.

## 4.2 Find the Gödel sentence $s_{Gödel}$ using diagonal proof

The previous chapter proved that some objects in the space of tensor-form systems have no corresponding interpreter, and the proof itself is complete and valid. The next thing we want to do is explore the internal structure of this object and reconstruct the object that leads to the incompleteness. This step, analogous to the original incompleteness theorem proving, is equivalent to finding out who the theorem is that causes the incompleteness.

To finish the proof of the Gödel Incompleteness Theorem for PAC Learnable Theory, we can find this certain sentence (well-constructed string) to make this proof more complete.

As the proof in the previous chapter, we prove that if

$$\exists B\ is\ the\ subspace\ of \{H \to H\},$$
$$\exists A\ is\ the\ subspace\ of\ H$$
$$s.t, span(B)span(A) = H$$

$$and\ dimB < dimH^2\ or\ dimA < dimH$$

$$\Rightarrow$$

$$\exists \boldsymbol{S_{Gödel}} = S_B \otimes S_A + S_{\neg B} \otimes S_{\neg A}$$
$$s.t, \nexists H_G \in H_{\otimes}\ satistfies\ H_G I = \boldsymbol{S_{Gödel}}$$
$$\Rightarrow$$
$$\exists s_{Gödel} \in S^*\ s.t, \nexists h \in H\ satistfies\ hI = \boldsymbol{s_{Gödel}}$$

Where $S^*$ is the closure of $S$, which means it could include some certain limit of $S$ components. In fact, $s_{Gödel} \in S^* and\ s_{Gödel} \notin S$ .

Since $span(B \otimes A) = H_{\otimes}$, $H_G$ could be decomposed as the following form:

$$H_G = \sum_{i=1}^{\dim B} \sum_{j=1}^{\dim A} \lambda_{ij} B_0^{(i)} \otimes A_0^{(j)}$$

Where $B_0^{(i)}$, $A_0^{(i)}$ indicate the eigenvector corresponding to $\{H \rightarrow H\}$ ,$H$， i is used as the i-th eigenvector to characterize axiom i. Given such a decomposition, a new object can be constructed in the following way：

$$\boldsymbol{s_{G\ddot{o}del}} = \sum_{i=1}^{\dim B} \sum_{j=1}^{\dim A} \mu_{ij} s_{B_0^{(i)}} \otimes s_{A_0^{(j)}} \in S_{G\ddot{o}del} \qquad (3-8)$$

$$\mu_{ij} = \begin{cases} \lambda_{11}, & \lambda_{ij} \neq \lambda_{11} \\ \lambda_{22}, & \lambda_{ij} = \lambda_{11} \end{cases}$$

Here we assume $\lambda_{11} \neq \lambda_{22}$. If $\lambda_{11} = \lambda_{22}$, we just need to find another element $\lambda_{ij} B_0^{(i)} \otimes A_0^{(j)} \in H_G$ s.t, $\lambda_{ij} \neq \lambda_{11}$ to replace $\lambda_{22}$ in our argument. Because $B$ and $A$ has less dimension than the full space, we can always find two different $\lambda_{ij}$. So $\boldsymbol{s_{G\ddot{o}del}}$ could be found, hence:

$$\boldsymbol{s_{G\ddot{o}del}} \in S\ s.t, \nexists h \in H\ s.t., hI = \boldsymbol{s_{G\ddot{o}del}}$$

■

Next, we calculate the dimension of $\boldsymbol{S_{G\ddot{o}del}}$.

Let $n = \dim(H) \geq 2, HI = S$,we have:

$$\dim(\boldsymbol{S_{G\ddot{o}del}}) = 2^n - n - 1$$

Prove:

$n = \dim(H)$ means H has at most $n$ linearly independent eigen vectors, which means you can compose an entangled $\boldsymbol{S_{G\ddot{o}del}}$ by choosing k of n eigen vectors. Since $A, B, C, \ldots \in H \Rightarrow A \otimes B \otimes C \otimes \ldots \in H_{\otimes}$ , so if we compose an entangled $\boldsymbol{S_{G\ddot{o}del}}$ by choosing k of n eigen vectors, this $\boldsymbol{S}_{\boldsymbol{G\ddot{o}del}}^{\binom{n}{k}}$ is linearly dependent with the same k eigen vectors composed by another way.

Therefore, $\dim \boldsymbol{S}_{\boldsymbol{G\ddot{o}del}}^{\binom{n}{k}} = \binom{n}{k}$

Since 1 of n eigen vectors cannot form an entangled $\boldsymbol{S_{G\ddot{o}del}}$.

So

$$\dim(\boldsymbol{S}_{\boldsymbol{G}\ddot{o}\boldsymbol{del}}) = 2^n - \binom{n}{1} - \binom{n}{0} = 2^n - n - 1$$

Note that if $\dim(H) \leq 1$, $\dim(\boldsymbol{S}_{\boldsymbol{G}\ddot{o}\boldsymbol{del}}) = 0$.

■

We incidentally prove again that if the interpreter space dimension is equal to 1 or 0, then there is a one-to-one correspondence between the interpreter space and the formal system space, in other words, it is complete. This conclusion corresponds to the original Gödel's completeness theorem.

Things below are just something we could discuss about this formula philosophically.

N=2， $\dim(\boldsymbol{S}_{\boldsymbol{G}\ddot{o}\boldsymbol{del}}) = 1$

Correspondingly, there is only one class of Gödel incomplete propositions in first-order predicate logic.

N=3， $\dim(\boldsymbol{S}_{\boldsymbol{G}\ddot{o}\boldsymbol{del}}) = 4$

Corresponding to our three-dimensional space, we can understand the 4 space-time, but we do not live in the 4-dimensional space. The whole of four-dimensional space is incomprehensible to us, and we can only understand three-dimensional slices

N=4, $\dim(\boldsymbol{S}_{\boldsymbol{G}\ddot{o}\boldsymbol{del}}) = 11$

N=5， $\dim(\boldsymbol{S}_{\boldsymbol{G}\ddot{o}\boldsymbol{del}}) = 26$

In contemporary physics, string theory believes that space is ten- or eleven-dimensional space (the maximum dimension is eleven dimensions, see the theoretical value of M), and in addition to the known three dimensions, the remaining dimensions are all in quantum mechanical measurements (10 to the minus ten). This theory is only hypothetical, although it has strict theoretical basis but has not been proved by experiments. String theory has bigger questions.

It seems that spatial time is coordinated only if it has ten or twenty-six dimensions, rather than the usual four. According to Einstein's theory of general relativity, there are outside of us.

There is more space, because space-time is not absolute, but relative, space can be changed, so from the three-dimensional space-time we live in, derived from other spaces.

N=6， $\dim(\boldsymbol{S}_{\boldsymbol{G}\ddot{o}\boldsymbol{del}}) = 57$

O(n)-> $\dim(\boldsymbol{S}_{\boldsymbol{G}\ddot{o}\boldsymbol{del}})$ =O($2^n$)

The dimension of the understandable part of the infinite generation task is O(n), so the dimension of the incomprehensible part is $O(2^{n})$. (Understandable here means that PAC is learnable and has an interpreter; Incomprehensible means that PAC is learnable but has no interpreter)

## *5 Examples to apply general Gödel incompleteness theorem*

### 5.1 Cantor diagonal proof

Let $H$ be a proper way to construct a number in $\mathbb{R}$ by $\mathbb{N}$.

$H$ has at least two "linearly independent": nonrepeating infinite decimals and rational number $\mathbb{Q}$, which could be trivially proved by the arguments $\sqrt{2} \notin \mathbb{Q}$ and $\mathbb{N}\sim\mathbb{Q}$. Therefore, $\dim(H) \geq 2$.

Let:

$$I = \{0,1,2,3,4,5,6,7,8,9\}\ composes\ of\ the\ basic\ elements\ of\ \mathbb{N}$$

$$A = \{a|a \in H\ s.t, aI \in \mathbb{Q} = S_A\}$$

$$\neg A = \{a|a \in H\ s.t, aI \in (\mathbb{R} - \mathbb{Q}) = S_{\neg A}\}$$

$$B = \{b|b \in \{H \to H\}, a \in H\ s.t, baI \in (0,1)\sim\mathbb{R}\}$$

$$\neg B = \{b|b \in \{H \to H\}, a \in H\ s.t, baI\ \notin\ (0,1)\sim\mathbb{R}\}$$

We have ("all $s \in S$ can be constructed" is tried to get by $H$):

$$span(BA) = H$$

And:

$$\dim(A) < \dim(H)$$

So $S_{\boldsymbol{G}\ddot{o}\boldsymbol{del}}$ can be constructed, for example:

$$\boldsymbol{S}_{\boldsymbol{G}\ddot{o}\boldsymbol{del}} = S_B \otimes S_A + S_{\neg B} \otimes S_{\neg A}$$

And $s_{\boldsymbol{G}\ddot{o}\boldsymbol{del}}$ can be produced by diagonal method in (3.3), which by the definitions above is exactly the diagonal number that cannot be presented by $\mathbb{N}$.

Let $n \to \infty$, especially, let $n = \aleph_0$, then $\dim(\boldsymbol{S}_{\boldsymbol{G}\ddot{o}\boldsymbol{del}}) = 2^{\aleph_0} = \aleph_1$.

One of the greatest achievements of this formula is that it gives a way to discuss continuum hypothesis. It shows that a denumerable infinity $\aleph_0$ could construct $\aleph_1$ by allowing any types of limits. Just like the $\boldsymbol{s_{Gödel}}$ constructed above, we give any real number $a$ between 0 and 1 an infinite decomposition:

$$a = 0 + a_1 \times 10^{-1} + a_2 \times 10^{-2} + a_3 \times 10^{-3} + \cdots$$

Consider $a$ as an infinite sum of eigen vectors $10^{-i}$, and let $a_i \in \{0,1,\dots,9\}$. Then any real number in this interval could be regarded as an expansion of the original reasonable numbers $10^{-i}$ and $\{0,1,\dots,9\}$. So the $\aleph_1$ is the PAC learnable expanded space of $\aleph_0$ space while some of the numbers in $\aleph_1$ could not be found in $\aleph_0$. In fact, $2^{\aleph_0}$ just shows that the Lebesgue measure of $\aleph_0$ in $\aleph_1$ is zero.

### 5.2 Original Gödel Incompleteness theorem

Let $H$ be all the legal logical procedure in the First-order logical system.

$H$ has at least two "linearly independent": legal procedure in Zero-order logical system and legal procedure in First-order logical system. Therefore, $\dim(H) \geq 2$.

Let:

$$I = \{P1, P2, P3, P4, P5\} = \{axioms\ of\ Peano\ Arithematics\}$$

$$A = \{a | a \in H\ s.t, \quad \exists\lambda > 0, \quad aI = \lambda \|\psi\|^2\}$$

$$\neg A = \{a | a \in H\ s.t, \quad \exists\lambda < 0, \quad aI = \lambda \|\psi\|^2\}$$

By the way:

$$A_\Phi = \{a | a \in H\ s.t,\ \exists\lambda = 0,\ aI = \lambda \|\psi\|^2\} = \Phi$$

$$\Phi\mathrm{I} = \emptyset$$

ΦI is not well-defined or just doesn't contain any element in S(E).

Let A also satisfies:

$$A = span(\{axioms\ of\ \text{Zero} - \text{order logical system}\})$$

Let B be similar to A:

$$B = \{b | b \in \{H \to H\}, a \in A\ s.t, \quad \exists\lambda > 0, \quad baI = \lambda \|\psi\|^2\}$$

$$\neg B = \{b | b \in \{H \to H\}, a \in A\ s.t, \quad \exists\lambda < 0, \quad baI = \lambda \|\psi\|^2\}$$

We have ("all $s \in S$ can be constructed" is tried to get by $H$):

$$span(B)span(A) = H$$

And:

$$\dim(A) < \dim\,(H)$$

So $S_{\boldsymbol{G\ddot{o}del}}$ can be constructed, for example:

$$\boldsymbol{S_{G\ddot{o}del}} = S_B \otimes S_A + S_{\neg B} \otimes S_{\neg A}$$

The tensor space method is isomorphic to Gödel number method, which is developed to ask something about the formal system itself.

The meaning of $\boldsymbol{S_{G\ddot{o}del}}$ is " $S_B \otimes S_A \sim prove\ that\ is\ true$ " and " $S_{\neg B} \otimes S_{\neg A} \sim disprove\ that\ is\ wrong$" which indicates the completeness of S.

All proposition space $S_{ij} \in \boldsymbol{S_{G\ddot{o}del}}$ is a real space that is a subspace of S, in other words, $\forall s \in S_{ij} \in \boldsymbol{S_{G\ddot{o}del}}, s\ is\ true.$

But we can construct $s_{\boldsymbol{G\ddot{o}del}}$ by diagonal method which is true and in S, but has no $h \in H$ to prove.

Therefore, formal system S is incomplete. The Gödel incompleteness theorem I has been proved.

The Gödel incompleteness theorem II "the consistence of S cannot be proved in S" can in a similar way to construct $\boldsymbol{S_{G\ddot{o}del}}'$.

$$\boldsymbol{S_{G\ddot{o}del}}' = S_B \otimes S_{\neg A} + S_{\neg B} \otimes S_A$$

The meaning of $\boldsymbol{S_{G\ddot{o}del}}'$ is " $S_B \otimes S_{\neg A} \sim prove\ that\ is\ wrong$ " and " $S_{\neg B} \otimes S_A \sim prove\ that\ is\ false$" which indicates the consistence of S.

Arguing that corresponding $s_{\boldsymbol{G\ddot{o}del}}'$ is a false proposition in S but cannot be proved by H. Therefore, the Gödel incompleteness theorem II has been proved.

Note that for all set of axioms $I\ with\ |I| > 1$ and logical system higher or equal than First ( $dim(N - order\ logical\ system) = N + 1, N \in \mathbb{N}$ ), the corresponding formal system S is complete. While for all Zero-order logical system or for all set of axioms that only contains one axiom, the formal system is complete.

### 5.3 Turing's Halting Problem

Let H be a Turing machine that receives as input information (encoded in 0s and 1s) and a procedure (also encoded in 0s and 1s). The Turing machine outputs whether the computation halts or runs forever.

$H$ has at least two "linearly independent": the process of information input and the process of procedure input. Therefore, $\dim(H) \geq 2$.

Let:

$$I = \{0,1\}$$

$$A = \{a|a \in H\ s.t, aI = s\ \boldsymbol{halts}\} = \{a|a \in H\ s.t,\ \exists\lambda > 0,\ aI = \lambda\|\psi\|^2\}$$

$$i.e., AI = \{S_A\ that\ halts\}$$

$$\neg A = \{a|a \in H\ s.t, aI = s\ \boldsymbol{never\ halts}\} = \{a|a \in H\ s.t,\ \exists\lambda < 0,\ aI = \lambda\|\psi\|^2\}$$

$$i.e., \neg AI = \{S_A\ that\ never\ halts\}$$

By the way:

$$A_\Phi = \{a|a \in H\ s.t,\ \exists\lambda = 0,\ aI = \lambda\|\psi\|^2\} = \Phi$$

$$\Phi\mathrm{I} = \emptyset$$

ΦI is not well-defined input or just doesn't contain any 0 or 1 in S(E).

$$B = \{b|b \in \{H \to H\}, a \in A\ s.t, \qquad \exists\lambda < 0, \qquad baI = \lambda\|\psi\|^2\}$$

$$i.e., B = \{if\ input\ is\ \text{halt}, then\ output\ is\ never\ halt\}$$

$$\neg B = \{b|b \in \{H \to H\}, a \in A\ s.t, \qquad \exists\lambda > 0, \qquad baI = \lambda\|\psi\|^2\}$$

$$i.e., \neg\ B = \{if\ input\ is\ \text{never halt}, then\ output\ is\ \ halt\}$$

We have ("all $s \in S$ can be identified to be halt or never halt" is tried to get by $H$):

$$span(BA) = H$$

And since $H$ contains both information input process and procedure input process, so:

$$\dim(A) < \dim(H)$$

So $S_{\boldsymbol{Gödel}}$ can be constructed, for example:

$$\boldsymbol{S_{Gödel}} = S_B \otimes S_A + S_{\neg B} \otimes S_{\neg A}$$

The tensor space method is isomorphic to 0-1 number input method, which is developed to ask something about the Turing machine itself.

The meaning of $\boldsymbol{S_{Gödel}}$ is " $S_B \otimes S_A \sim let\ the\ halt\ ones\ never\ halt$ " and " $S_{\neg B} \otimes S_{\neg A} \sim let\ the\ never\ halt\ ones\ halt$" which indicates the decidability of S.

We can get the corresponding $\boldsymbol{s_{Gödel}}$ which cannot be decided halt or not in any circumstances on the hypotheses in this part. Therefore, Turing machine $S$ is not decidable.

By analogy to Gödel Incompleteness theorem II, there is a trivial conclusion:

The consistence cannot be proved within in S.

### 5.4 EPR Paradox and Quantum Entanglement

Firstly, we will describe a simplified version of the EPR paradox first introduced by David Bohm.[31](*31*)

Consider the decay of the neutral pi meson into an electron and a positron

$$\pi^0 \rightarrow e^- + e^+$$

Assuming the pion was at rest, the electron and positron fly off in opposite directions. Now, the pion has spin zero, so conservation of angular momentum requires that the electron and positron are in the singlet configuration:

$$\frac{1}{\sqrt{2}}(\uparrow_-\downarrow_+ - \downarrow_-\uparrow_+)$$

Now if you measure the spin of the electron, say you get spin up, immediately you know that someone 20 meters (or 20 light years) away will get spin down, if that person examines the positron.

Next, let's consider a more complex version:

Consider a two-level system H, $|\phi_a\rangle and\ |\phi_b\rangle$ are two subspaces of H, with $\langle\phi_i|\phi_j\rangle = \delta_{ij}$, (for example, as $|\phi_a\rangle$ may represent spin up and $|\phi_b\rangle$ spin down). The two-particle state:

$$\alpha|\phi_a(1)\rangle \otimes |\phi_b(2)\rangle + \beta|\phi_a(2)\rangle \otimes |\phi_b(1)\rangle$$

Which cannot be represented by

$$|\psi_r(1)\rangle \otimes |\psi_s(2)\rangle$$

[The proof of this is persistent with the argument in $S_{Gödel}$]

So far the fundamental completeness of this quantum system hasn't been broken, since each measure still exists for the entangled state. But remember what Gödel had said to us: There is a $s_{Gödel}$ state which is an entangled state but cannot be measured at all, i.e., $s_{Gödel}$ state has a certain possibility to be the first tensor state or the second, but it just cannot be detected to collapse its own wave function. This phenomenon is isomorphic to the Gödel incompleteness theorem.

For one particle spins on two different directions composing a Hilbert space $H$, one version of $S_{Gödel}$ could be written into the following form:

$$S_{Gödel} = \uparrow\otimes\uparrow + \downarrow\otimes\downarrow$$

The corresponding $s_{Gödel}$ will destroy the completeness of $H$.

And the other $S_{Gödel}{}'$ could be written into the following form:

$$S_{Gödel}{}' = \uparrow\otimes\downarrow + \downarrow\otimes\uparrow$$

The corresponding $s_{Gödel}{}'$ will indicate the consistence of $H$ cannot be proved within $H$.

But if $\dim(H) = 1$, all the paradoxes above will disappear, which is when you put the particles through the Stern-Gerlach box, or say, the collapse of the wave function in quantum mechanics and quantum locking.

## *6 Discussion about Gödel incompleteness space*

### 6.1 Syntactic duality completeness leads to semantic logical duality incompleteness

Another interpretation of the generalized Gödel incompleteness theorem is that if two nontrivial systems are syntactic dual and complete to each other, then their semantic and logical duality is incomplete. Of course, it should not be too rudimentary. 0-dimensional and 1-dimensional are complete, but those above 2-dimensional are not. Therefore, 0-dimensional and 1-dimensional systems are called trivial systems.

In a trivial system, all objects have only one type of reflexivity. The entire ordinary system has only one type of self-reference or no self-reference at all. A 0-dimensional system will be directly equivalent to a simple listing of facts or a set with no internal structure at all. Therefore, we can say that the 0-dimensional system is equivalent to the meaning of "experience". Starting from the first dimension, the chain structure from axioms to theorems becomes apparent. However, with only one dimension, there is no coherent structure of mutual entanglement, and thus it is also simple and complete, corresponding to a 0-order predicate logic system. Starting from two dimensions, that is, from the equivalent of first-order predicate logic, Gödel's incompleteness theorem will emerge in the system. The n-dimension (where n is a natural number greater than or equal to 1) describes obtaining the entire proposition tree starting from the "a priori" axiom and through the "a priori" logical derivation pattern. Therefore, we call systems above one dimension the systems that describe "a priori".

For a system where each object's automorphism symmetry group contains $\mathbb{Z}^2$ fundamental group logically (that is, for any object within the system, there are two non-equivalent construction loops that can return to themselves), if another system's syntactic translation of it is complete, then its semantic logical translation (in the sense of allowing "limits" of Cantor diagonal proof) is incomplete. This proposition is an obvious sufficient condition for the dimensional version of the generalized Gödel incompleteness theorem. Because if each propositional object can return to itself through some kind of construction loop, then this kind of construction loop corresponds to a construction axiom. If each propositional object has at least two non-equivalent loops, then the system contains at least two non-equivalent construction axioms, that is, the dimension of the system is greater than or equal to 2. Thus, we can formally transform the study of the logical structure of a system into the study of the topological space in which the system logic is located.

Therefore, Gödel's incompleteness theorem does not stem from the robustness of the set of natural numbers but from the duality between the numbers constructed by Gödel's ligands and logic. Therefore, the only axiom of a defective proposition in this dual space constructed by Gödel is the original Gödel proposition, and apart from this proposition, there are no other kinds of Gödel incomplete propositions as shown in Fig. S8. This assertion stems from the fact that the dimension of first-order predicate logic is 2, and the corresponding dimension of the Gödel space is $2^2 - 2 - 1 = 1$. The robustness of the set of natural numbers only guarantees the semantic completeness of duality.

For HI=S, the formal information S is obtained by observing I through H. H and S are two categories. H is constructive (axiom + logical constructiveness, reasonable is existence), and S is free (well-structured constructiveness, existence is reasonable). If there exists a faithful functor f from H to S and the initial dimension of H's construction is greater than or equal to 2, then f must be incomplete (not surjective). Interestingly, semantic duality is guaranteed by the "limit" construction method allowed by H, which in turn leads to the incompleteness of H itself and its extensibility, just like the extension from rational numbers to real numbers, H will be extended to $H^*$.

Sets and free groups are naturally associated with each other. Then, ZFC can construct incomplete theorems even when the free group language allows it externally, and such incomplete theorems will not occur within the system under the ZFC construction framework.

This characterizes that many incompleteness avoided by mathematicians through constructiveness in the past will naturally become incompleteness due to the external free language. The limitations of the cognitive capability are shown in Fig. S9.

### 6.2 First-order predicate logic is incomplete for $\mathbb{N}$, but it is complete for $\mathbb{Q}(\sqrt{2})$

First-order predicate logic contains two inequivalence loops, which are equivalent to two inequivalence axioms. The number of incomplete spatial dimensions it can span is $2^2 - 2 - 1 = 1$ .Therefore, it is only necessary to supplement a class of numbers that can be represented in the limit sense of rational numbers but are not contained within rational numbers as their mapping to incomplete theorems to expand it into a complete space. For example, $\sqrt{2}$ is supplemented as a mapping of incomplete theorems, so first-order predicate logic is relative to$\mathbb{Q}(\sqrt{2})$ is complete.

Thus, for n-order predicate logic to be expressed completely, it needs to be extended over the field of rational numbers $2^n - n - 1$ irrational numbers that are orthogonal in the sense of number theory. Meanwhile, if $n \to \aleph_0$, then $\aleph_1$, that is, the set of real numbers $\mathbb{R}$, is needed to be complete. Therefore, for any countable infinite-order predicate logic, it can be completely expressed by $\mathbb{R}$.

### 6.3 Dimension correspondence

The family that understands the full trajectory of an object's movement requires additional dimensions.

Because the process of motion actually follows the situation described by Zeno's paradox: that is, if our motion process is decomposed into infinite steps, the infinite process itself is inoperable. However, in fact, we can move.

This is one of the most classic limit logic constructions.

From this, we regard the concept of "motion", especially the motion on continuous sets, as a situation described by an incompleteness space with discrete operation steps. In particular, quantum mechanics is a discrete space, and the motion in its corresponding incompleteness space can be continuous.

Starting from the dimension calculation formula of Gödel space, understanding the motion of a 2-dimensional object requires 1 dimension, understanding the motion of a 3-dimensional object requires 4 dimensions, understanding the motion of a 4-dimensional object requires 11 dimensions, and understanding the motion of a 5-dimensional object requires 26 dimensions.

Starting from three-dimensional space, physical objects are defined as a family of trajectories, and the negative meaning of physical objects is defined as all other possible trajectories of other families. The concept of this family of trajectories comes from historical summation, and the definition of "no" is equivalent to the definitions of 0, 1, or spin above and below. When the spatial dimensions of 3, x, and y are viewed separately, the basis axiom that can be mutually transformed (rotating coordinate systems) and orthogonal emerges! The axiomatic system thus developed requires a supplementary dimension to fully understand the meaning of its motion (in the sense of limits), which is the dimension of Gödel's incompleteness theorem. From this, we can hold that if a discrete quantum system in three-dimensional space is taken as the basis, the four dimensions of the Gödel incompleteness space it can span can exactly correspond to: the continuous X-axis, the continuous Y-axis, the continuous z-axis, and the dimensions corresponding to $x \otimes y \otimes z$. The dimension $x \otimes y \otimes z$ should describe a kind of variation over the entire domain, that is, under this dimension, motion becomes possible. This dimension corresponds most appropriately to the continuous "time" dimension. A more appropriate way to describe the time dimension here is: in the case of three continuous axes, if you want to describe "motion", the most suitable supplementary construction of "time".

For example, we hope that: $x \otimes y$ corresponds to a continuous axis. For instance, if a countable number of bases on the X-axis reach a certain irrational number through the construction of the Cauchy sequence, we will map this irrational number to a certain ground state of y. This is equivalent to being able to construct the correspondence of each unreasonable decimal point on the Y-axis. Therefore, two discrete axes are mapped to a continuous axis in the sense of logical limits. There are two extensions of three types, so there are three continuous axes.

From this perspective, the contradiction between quantum mechanics and general relativity is obvious. The former is the basis for describing systems, while the latter is the result of continuous motion in its limit sense. Conclusions drawn under general relativity should have no corresponding interpreter in the quantum mechanics system. That is to say, Within the discrete framework of naive quantum mechanics, no corresponding physical objects that can describe continuous space and continuous time can be found. For example, you can never continuously observe the continuous motion trajectory of a quantum.

However, at the same time, this idea precisely reconciles the contradiction between quantum mechanics and general relativity, because in the sense of limits, quantum mechanics can logically reach general relativity, but conversely, it cannot find a counterpart. That is to say, the quantum or graviton corresponding to the continuous space should have no corresponding interpreter within the framework of quantum mechanics, that is, no corresponding quantum can be found. This does not mean that gravitons do not exist. On the contrary, they should exist (be a true proposition) and be unobtainable (no corresponding interpreter can be found).

From the perspective of p: "Every proposition A is accompanied by an Aut(A) group containing the fundamental group $\mathbb{Z}^2$, then the distribution of the construction of incomplete theorems throughout the formal system space is consistently continuous (equivalent to the existence of a pair of orthogonal axioms that can consistently act on each proposition). Then a sufficient condition for the existence of Gödel incompleteness in the system is p. Such consideration is more convenient because there is no need to clean up the entire axiomatic system anymore, but only a review of the number of non-equivalent automorphisms for any single research object is required.

Therefore, as long as the inequivalent automorphisms of the objects studied in string theory are five and those of superstrings are four, then the correspondence between Gödel's incompleteness space and the dimensions of string theory must be correct.

We put forward such a conjecture: The five bosons that string theory can describe have structured the string theory system into five non-equivalent isomorphisms (gluons, photons, W bosons, Z bosons, Higgs bosons). Here, the influence of boson color is ignored and bosons are classified into five categories.

From the study of phonons, we can see that bosons precisely correspond to the symmetry breaking of the system. A symmetry breaking will cause the system to fall into a certain state. Asymmetric (physically speaking, it should be that they cannot be naturally transformed into each other) symmetry breaking will cause the system to fall into an asymmetric phase. Therefore, if the five bosons precisely correspond to five unequal symmetry breaks, and the colors described by QCD precisely enable the same bosons to transform into each other, then the argument of this pair of string theory that a 26-dimensional space is a Gödel incompleteness space with a base of 5 dimensions will be valid.

A symmetry breaking corresponds to a ground state, a state that falls in. However, the existence of a symmetry breaking means that there must be at least two states, the critical state before the fall and the state after the fall. However, if there are only two states, then these two states will be mutually equivalent, which means there are no incomplete states in the entire model. This exactly corresponds to the situation where $2^1 - 1 - 1 = 0$. Therefore, at least two symmetry breaks, one falling to the left and the other to the right of the mountain, correspond to two ground states. Coupled with the fact that the mountain top itself serves as a critical state, this is the simplest state with an incomplete phenomenon, that is, the critical state.

From here we can see that for a physical system, if its non-equivalent phases are regarded as orthogonal ground states, then the incompleteness space it spans corresponds to all the entangled states. When all phases of an entangled state occur in the same space and the entire space has the potential to fall into each phase, it is precisely the "critical state" described by statistical mechanics. It can only be described statistically. Observing and recording the particle states one by one makes little sense, once again confirming that this is a state where no corresponding interpreter can be found. Conversely, if we start from the axiomatic basis of the original system, we can only understand the critical state in a statistical sense, and only understand the incompleteness space in a statistical sense (being able to handle its distribution and analyze its dimensions as a whole), and will never be able to directly find the corresponding interpreter or directly understand the critical state. Only methods such as order parameters can be used to describe the critical state as a whole.

For superstring theory, it semantically describes the four fundamental forces: the weak interaction force, the strong interaction force, the electromagnetic force and the gravitational force. Each force corresponds to a type of phase produced by a kind of symmetry breaking that is orthogonal to each other. Just as the weak interaction force and the electromagnetic force are unified (or entangled) at high energy, it is precisely because of symmetry breaking that they split into different forces at low energy. Then, to describe all phenomena logically and completely, the theory needs to be placed in an 11-dimensional space.

But unfortunately, if this kind of correspondence is rigorous, then superstring theory is doomed to be unable to find a corresponding interpreter, that is, any phenomenon in the 11-dimensional space cannot be directly observed by experiments. But fortunately, this also indicates that for such a theory that can describe the four fundamental forces and can only be fully described in an 11-dimensional space, it is a completely correct true proposition, and no direct experimental proof is needed and cannot be found.

This is as if the entangled state is real, but we can never directly observe it. We can only experimentally measure it in a hypothesis-deductive manner time and again. However, none of these are direct evidence of the existence of entangled states, because for systems with electron spins of up and down as the base, entangled states are their incompleteness spaces, where corresponding interpreters cannot be found. Meanwhile, the emergence of entangled states as incompleteness spaces of critical states precisely proves that their existence is real and cannot be observed.

Thus, although the above discussion may be rather lacking in details, if it is correct in principle, we have to accept the fact that:

For string theory and superstring theory, on the one hand, it has been proved that they are both correct and true, and can never be directly observed by experiments.

From now on, fundamental theoretical physics (specifically string theory and superstrings) may enter a new paradigm where theories are true, the phenomena corresponding to the theories are all real, but can never be directly observed. Thus, physics broke away from the paradigm that had to be experimentally verified, because the theory fully interpreted the phenomenon, but the theory itself was in an incompleteness space that could not be directly observed.

It should be emphasized that if fundamental theoretical physics (specifically string theory and superstrings) is truly in a space that is true but incomplete, then only the experimental pursuit of fundamental theoretical physics fails, while other macroscopic or empirical experiments remain and are of great significance. Conversely, the deductive experiments of hypotheses in the statistical sense given by theoretical physics still make sense, but they can never directly prove or observe the existence of strings.

*Electromagnetism*

N=2， $\dim(\boldsymbol{S}_{\boldsymbol{G}\ddot{o}\boldsymbol{del}}) = 1$

Correspondingly, given axioms electric field E and magnetic field B, there will be one certain type of incompleteness axiom that look like this:

$$S = E \otimes B + \neg E \otimes \neg B$$

This corresponds to the synchronization of the electric and magnetic fields of electromagnetic waves at the same moment, predicting the characteristics of the quantum mechanical wave-particle duality and the anti-Bell inequality of electromagnetic waves. This ¬ could be rotating 90 degrees so that $\neg E\ and \neg B$ is orthogonal to $E\ and\ B$ relatively.

*Relativity*

N=3， $\dim(\boldsymbol{S}_{\boldsymbol{G}\ddot{o}\boldsymbol{del}}) = 4$

For the Lorentz transformation group SO (1,3), it can be divided into the direct sum of two parts SU (2). The fundamental group of SO (1,3) is $\mathbb{Z} / 2\mathbb{Z} = \{\pm 1\}$, thus it contains two loops. Translation transformation contains a loop. Therefore, the Poincaré transformation group as a whole is equivalent to three loops - three axioms. Therefore, the incompleteness space formed by its entanglement has four dimensions.

*Superstring theory*

N=4, $\dim(\boldsymbol{S}_{\boldsymbol{G}\ddot{o}\boldsymbol{del}}) = 11$

On the one hand, Poincaré transformation group plus $U(1) \times SU\ (2) \times SU\ (3)$ is the symmetry of the four fundamental forces. Among $U(1) \times SU\ (2) \times SU\ (3)$, only the fundamental group $U(1)$ has 1-dimension non-trivial fundamental group $\mathbb{Z}$, so the sum is 4-dimension axioms forming 11-dimension incompleteness space.

On the other hand, the four fundamental forces require the force to fall into four phases. If the number of phases is equal to the dimension of the base space, at least 11 dimensions are needed. If there are four major types of gauge bosons, then there are four major categories of symmetry breaking, and a complete description of their nontrivial combinations requires 11 dimensions.

*Boson String theory*

N=5， $\dim(\boldsymbol{S}_{\boldsymbol{G}\ddot{o}\boldsymbol{del}}) = 26$

Poincaré invariance (3-dimension axioms) plus Diff invariance (1-dimension axiom) plus Weyl invariance (1-dimension axiom) capture the symmetry of boson string theory. So the whole incompleteness space of this theory is 26 dimensions, which corresponds to its 26-dimension space to make boson string theory make sense in physics.

For boson string theory and superstring theory, on the one hand, it has been proved that they are both correct and true, and can never be directly observed by experiments.

From now on, fundamental theoretical physics may enter a new paradigm where theories are true, the phenomena corresponding to the theories are all real, but can never be directly observed. Thus, physics broke away from the paradigm that had to be experimentally verified, because the theory fully interpreted the phenomenon, but the theory itself was in an incompleteness space that could not be directly observed.

It should be emphasized that if fundamental theoretical physics (specifically string theory and superstrings) is in a space that is true but incomplete, then only the experimental pursuit of fundamental theoretical physics fails, while other macroscopic or empirical experiments remain and are of great significance. Conversely, the deductive experiments of hypotheses in the statistical sense given by theoretical physics still make sense, but they can never directly prove or observe the existence of strings.

*Ads/CFT*

For dimensions 2 to 1

Electromagnetic waves are quantized for electric and magnetic field systems, featuring entanglement, uncertainty, violation of Bell's inequality and many other quantum phenomena. Directly adding the entangled state dimension is equivalent to the overall phenomenon becoming classical, that is, using the entangled state to measure the entangled state to obtain the classical phenomenon

For dimensions 3 to 4

A three-dimensional quantum base vector can also grow into a four-dimensional space, and quantum phenomena transform into classical theories that are one dimension higher. Specifically, the time variation transforms from a completely discrete single dimension to a dimension of the overall system picture, entangling with other dimensions in the form of Venkovski or the spatial form of Ads, thereby encompassing the originally incomplete quantum phenomena in a certain direction (such as the most concerned gravitational direction), and thus the Ads/CFT duality is natural.

For dimensions 4 to 11

If the quantum gravity theory is to be classicized, only one dimension needs to be supplemented. This precisely proves that in the process of entangling a 4-dimensional base space into an 11-dimensional one, the pure classical gravity only occupies one basic entangled state.

For dimensions 5 to 26

The n+ 1-dimensional and n-dimensional CFT duality of string theory in the classical sense proves that the appearance of string theory in a classical way can merely be a phenomenon produced by supplementing one-dimensional entangled states to the axiomatic system based on CFT. Therefore, classical string theory itself can only appear as a single-degree-of-freedom entangled state of a CFT.

### 6.4 quantum observability

The following is to prove that for the observation of quantum objects (objects that can be described in matrix mechanics and its equivalent forms), PAC can learn ⇔ observable, while the quantum superposition state ⇔PAC cannot learn ⇔ unobservable.

The starting point of the following discussion comes from: According to the theory of decoherence, a single observation of a particle yields a single result, which is equivalent to using a backup set with one element in this case. Thus, the only learnable distribution of PAC is the δ distribution.

Firstly, allow me to rewrite the equivalent equation as follows:

(1) Information equivalent Eq.

$$\hat{A}|I\rangle = |s_A\rangle = \lambda_A|\psi\rangle\langle\psi|1\rangle = |\psi\rangle\langle\psi|\lambda_A\rangle$$

(2) Energy equivalent Eq.

$$\left|\hat{A}|I\rangle\right|^2 = \langle I|\widehat{A^\dagger}\hat{A}|I\rangle = \langle s_A|s_A\rangle = |\lambda_A|^2\langle 1|\psi\rangle\langle\psi|1\rangle = \langle\lambda_A|\psi\rangle\langle\psi|\lambda_A\rangle$$

(3) Observation equivalent Eq.

$$\hat{A}|I\rangle = |s_A\rangle = \lambda_A|\phi_A\rangle\langle\phi_A|1\rangle = \widehat{H}|\phi_A\rangle\langle\phi_A|1\rangle$$

Or

$$\hat{A}I = s_A = \lambda_A|\phi_A|^2 = \phi_A^*\,\widehat{H}\phi_A$$

$\widehat{H}$ is an operator on $\phi_A$ which is an eigen function of $\widehat{H}$ with eigen value $\lambda_A$.

For a certain wave function $\psi$ which is not an eigen function of $\widehat{H}$ but the superposition of some eigen functions, then

$$\nexists\lambda \in \mathbb{R},\ \widehat{H}\psi = \lambda\psi$$

When we observe $\psi$ with $\hat{A}$ to find quantity $H$, then

$$\hat{A}|I\rangle = |s_A\rangle = \lambda_A|\phi_A\rangle\langle\phi_A|1\rangle = \widehat{H}|\phi_A\rangle\langle\phi_A|1\rangle$$

Where $\phi_A$ is one of the eigen functions of $\widehat{H}$ with eigen value $\lambda_A$.

If one observes a superposition $\psi$ which is not an eigen function:

$$\widehat{H}|\phi_A\rangle\langle\phi_A|1\rangle \neq \lambda|\phi_A\rangle\langle\phi_A|1\rangle$$

We say that the quantity $H$ of $\psi$ is not PAC learnable, in other words, we cannot observe it in $O(\mathbb{P}(t))$ time.

PAC learnable observation of quantity $H$:

$$\hat{A}\psi = s_A = \lambda_A|\phi_A|^2 = \phi_A^*\,\widehat{H}\phi_A$$

PAC un-learnable observation:

$$\nexists \lambda_A, s.t.\ \phi_A^* \hat{H} \phi_A = \lambda_A |\phi_A|^2$$

One observation process of $\psi$ of quantity $H$, only one outcome would show up.

$$\Delta x \Delta p \geq \frac{\hbar}{2}$$

Once we measure the position $x$ at one time, the possibility distribution of momentum $p$ is spread out for $\Delta x \ \rightarrow 0,\ \ \Delta p \rightarrow \infty$, and vice versa.

If you are determined to examine $x$ and $p$ at the same time, the Fourier transform relationship between $\psi(x)\ and\ \tilde{\psi}(k)$ will collapse. The very use of our Fourier transform in quantum mechanics would be destroyed (which is really fundamentally bad), or if we consist to say Fourier transform has no flaw, then we can say "examine $x$ and $p$ at the same time" is PAC unlearnable, which means we cannot observe in $O(\mathbb{P}(t))$ polynomial time, or just say we cannot possibly understand this phenomenon, this outcome result.

Understanding the family of the full trajectory of an object's movement requires additional dimensions. The families of all trajectories are combined to form a critical state. The collapse condition of this critical state is what quantum mechanics calls "observation".

For double-slit interference, without observation, the superposition state formed by the combination of all possible paths taken by particles is an incompleteness space spanned by various orthogonal possible base paths.

The observation effect will cause states that originally cannot find a PAC-learnable interpreter within the system to have to show you a PAC-learnable phase (a certain eigenstate). Therefore, originally, what was received on the screen was the particle positions displayed in the completely incompleteness space. After observation, only the paths formed by individual particles through a single slit can be shown.

### 6.5 The critical state is the Gödel incompleteness space

The critical state is a Gödel incompleteness space, the number of phases is the number of dimensions of the original space, and the number of order parameters required to describe the critical state completely is the number of dimensions of the Gödel incompleteness space

The complete description of any critical state requires the number of dimensions of the Gödel incompleteness space. Any boson corresponds to a kind of symmetry breaking. Symmetry breaking causes the system to fall into the axiom number corresponding to the base space.

If the overall phase (such as the upward and downward spin of the ferromagnetic overall) is taken as the basis, the incompleteness space as a whole can be understood by several parameters of the incompleteness space dimension (such as temperature). However, if we look at the ferromagnetic particles one by one, the computational complexity of the number of parameters required for the system formed by the entanglement of n particles as a whole is O($2^n$), which is a PAC unlearnable problem. It cannot be observed in the polynomial sense, that is, it cannot be observed.

Starting from the incompleteness space, we can re-obtain the long-range correlation, scale invariance and power-law distribution phenomena of the critical state.

1. Long-range correlation

For a system composed of n particles, consider the incompleteness space formed by the entanglement of its various states. Then the formation of this entangled state is independent of the absolute position of the particle space. For two particles at any position, entangled states can be found in an incompleteness space. Therefore, one particle and another particle can exhibit long-range correlation.

2. Scale invariance

For n particles of a very large order of magnitude, the state of the overall composition of the incompleteness space they form is denoted as S. If n particles are regrouped and k particles are taken as a group for renormalization, and there is $k \ll n$. Then for the regrouped particles, approximately n/k particles of a very large order of magnitude will be entangled with each other, also forming a set of states S' in an incompleteness space. If the selection of the ground state observed remains unchanged and $n \approx n/k$, then S and S' are similar in the sense of various statistical properties.

3. Power-law distribution

If a certain statistical distribution of a family of particles has long-range correlation (long-tail effect) and scale invariance, then this distribution must be a power-law distribution.

The self-organizing critical theory holds that systems composed of a large number of interacting components will naturally develop towards the self-organizing critical state. When the system reaches such a state, even a very small interference event may cause a series of disasters in the system.

Very small interfering events can be described by the concept of fluctuations in physics.

Then we can obtain such a view:

The components of the system form various entangled states through a large number of interactions. All these entangled states are in the space described by the incomplete theorem.

When the system reaches such a state, the process of perturbation-induced catastrophe is equivalent to falling from the entangled state to the eigenstate, and from the incompleteness space to the ground state (δ distribution) that PAC can learn under a single measurement.

Therefore, the specific direction of the rise and fall (which state it will break into) is unlearnable for PAC.

Examples:

1. Ferromagnetic phase transformation

There are two ground states of ferromagnetic phase transitions: all spin upward and all spin downward shown in Fig. S10.

Its incomplete spatial dimension is $2^2 - 2 - 1$=1. Therefore, to describe the state of the critical state, a one-dimensional parameter needs to be supplemented. This is the order parameter we commonly use in the Ising model: the Hamiltonian. Or it can also be expressed more simply by temperature.

When the temperature rises, the particles become more disordered and change from ferromagnetic to paramagnetic.

When the temperature drops, the particles become more ordered and change from paramagnetic to ferromagnetic.

Because we divide the states into two ferromagnetic states and the paramagnetic state of the critical state from the perspective of macroscopic globalism, the specific change process of the specific spin of each particle predicted by the interaction mechanism at the microscopic level is unpredictable. However, on the whole, we can understand that the supplementation of an order parameter can explain the occurrence of this phase transition process, although paramagnetism is incomplete compared to the two ground states of ferromagnetism.

From the perspective of non-equivalent loops, there is no such path that does not pass through the critical state, which enables us to change from all spins downward to all spins upward. This reflects that the state transition is historically untraceable, that is, it embodies the idea that "operations that are reversible at the micro level are irreversible at the macro level".

In addition, the T drop that becomes all upward and the T drop that becomes all downward are not equivalent. Moreover, we only have one observable macroscopic quantity, T, so we cannot distinguish which phase this T decline will fall into.

2. the three-phase transition of water：

Taking solid-liquid-gas (S, L, G) as the three ground states, the incompleteness spaces formed include solid-liquid coexistence (S⨂L), liquid-gas coexistence (L⨂G), solid-gas coexistence (S⨂G), and the triple point (S⨂L⨂G). Combining all the phases yields the classic three-phase diagram (excluding supercritical water). The above discussions are shown in Fig. S11.

Its incomplete spatial dimension is 2^3-3-1=4. To completely describe the solid-liquid-gas three-phase states of a family of water particles, four parameters are required. This corresponds to pressure P, volume V, temperature T and latent heat of phase transition L. It follows the Clabron equation in the equilibrium state：

$$\frac{\mathrm{d}p}{\mathrm{d}T} = \frac{L}{T\Delta V}$$

At a certain ground state, the dimensions required for this description will degenerate. For example, for an ideal gas, PV=nRT.

The transformation between any two states can be characterized by changing only one order parameter given other order parameters.

The environmental conditions required for water to remain in a supercritical state are extremely demanding, with air pressure and temperature playing a dominant role. This is because only when both air pressure and temperature precisely reach a very appropriate proportion will water transition to a supercritical state under the combined influence of pressure and temperature. At this point, new arrangements of molecules and atoms in water emerge. So, water in the supercritical state is in a superimposed state of semi-gaseous and semi-liquid. At this time, the densities of gaseous water and liquid water are the same. It is precisely because this supercritical superimposed state is as confusing as the half-dead and half-alive cat in Schrodinger's ideal cat experiment.

So we can see that water in the supercritical state is the expansion of the entangled state "liquid phase ⊗ gas phase" (L⊗G) dimension in an incompleteness space, and it also conforms to the characteristics of the superposition state.

3. the predator-prey model:

For the simplest predator-prey model, its ground state can be regarded as three types shown in Fig. S12: (1) The predator eliminates the prey; (2) Predators become extinct and prey become rampant; (3) Steady-state symbiosis, where the populations of the two species rise and fall periodically.

The system of equations required to describe such a process:

$$\frac{dx}{dt} = ax - bxy$$

$$\frac{dy}{dt} = -cy + dxy$$

$x$ represents the prey, y represents the predator, t represents the time step, a represents the natural fertility rate of the prey, b represents the energy conversion rate consumed by the prey, c represents the natural mortality rate of the predator, and d represents the energy conversion rate at which the predator obtains energy from the prey to reproduce.

The free parameters required to describe such a three-phase system in such a complete way are exactly $2^3 - 1 - 1 = 4$.

4. Take the description of quantum mechanics using Hilbert space as an example:

Quantum mechanics that would fall into countable and infinite eigenstates would require Hilbert Spaces under real infinite eigenstates (continuous Spaces) to fully describe the "critical state" before collapse.

However, we can only observe the non-entangled eigenstates because entangled states do not have corresponding interpreters.

**6.6 Zero absolute temperature**

Consider such a family of n particles, whose two phases are rest and motion.

At the same time consider such an idea: absolute motion and relative rest.

Then each time we take a particle as the reference frame, the particle itself is equivalent to being stationary. Then the entire system S can be described as:

$$S_1 = \uparrow_1 \otimes \downarrow_2 \otimes \downarrow_3 \otimes \cdots$$
$$S_2 = \downarrow_1 \otimes \uparrow_2 \otimes \downarrow_3 \otimes \cdots$$
$$S_3 = \downarrow_1 \otimes \downarrow_2 \otimes \uparrow_3 \otimes \cdots$$

Among them, $\uparrow_k$ indicates that the KTH particle is at rest (selected as the reference frame), $S_k$ represents the description of S at this time, and various descriptions are equivalent under one coordinate transformation.

At this point, all these descriptions are complete descriptions of S. The computational complexity of the description class with the number of particles n is O(n). At this time, the incompleteness dimension is $2^1 - 1 - 1 = 0$, that is, the system is complete at this moment, and the corresponding bit is a system that can be stationary.

Consider reducing the speed of the particles to rest one by one. This is, in a sense, achievable for operating on each particle one by one.

Each time a particle is reduced to the velocity at rest relative to the particles in the reference frame, it is equivalent to at a certain bit k, and the system becomes capable of taking $\uparrow_k$ again. This is equivalent to adding a bit of basis, thereby generating an entangled state of $2^2 - 2 - 1 = 1$, corresponding to a certain two-bit system that can be stationary.

If we slow down the particles of this system one by one, attempting to gradually slow them down until they all come to rest relative to the particles in the reference frame, then the computational complexity of the number of dimensions of the states in the incompleteness space prepared in this process is $O(2^n)$.

That is to say, throughout the entire process of preparing absolute zero, it is necessary to conduct PAC-learnable observations on all possible entangled and incomplete states that may arise during the process. Such an observation process is bound to consume energy in the sense of information theory, thus forming a potential barrier that emerges when approaching absolute zero. Therefore, the height of this potential barrier should increase exponentially, corresponding to an exponential increase in possible incomplete states.

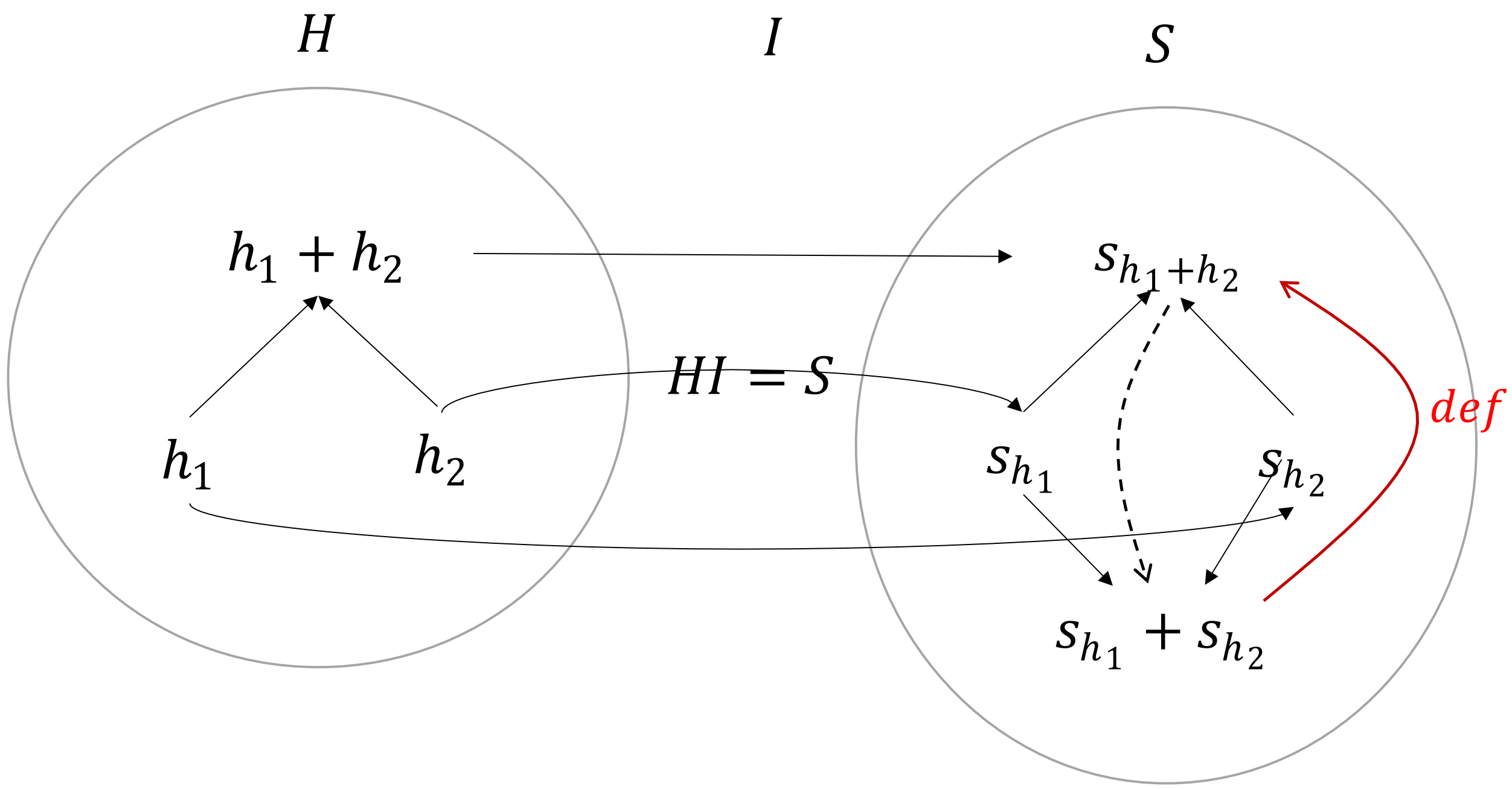

**Figure S1. the category diagram of the definition of the addition.** This commutive diagram between two categories is showing that the cognitive process $HI = S$ preserves the addition defined in interpreter space $H$ by addition of the corresponding distributions into cognitive formal space $S$, which is in line with the direct sum defined in category theory.

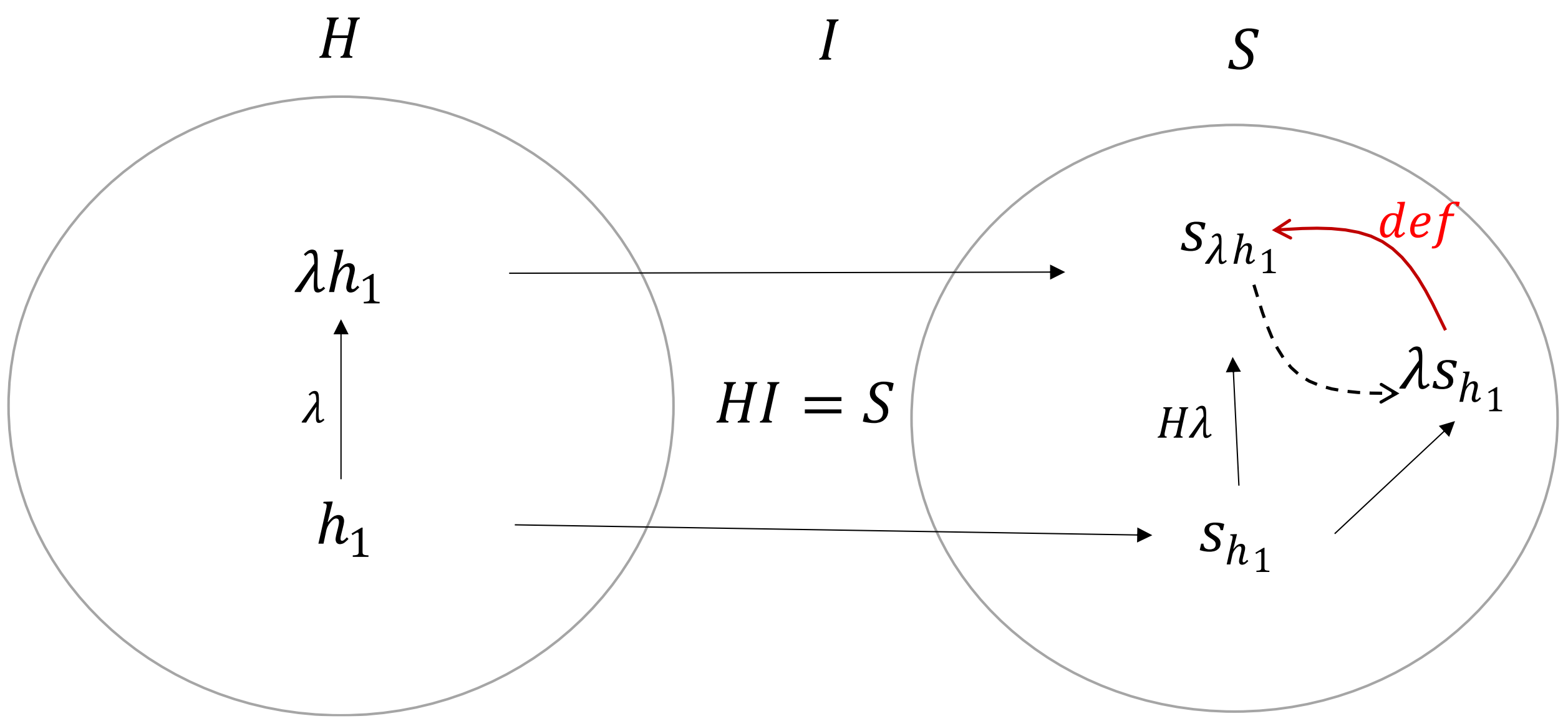

**Figure S2. the category diagram of the definition of the scaling.** $\lambda h_1$ is the direct sum $\lambda$ times of $h_1$ with $\lambda h_1 I = s_{\lambda h_1}$ and $h_1 I = s_1$. We define $\lambda s_{h_1}$ by defining the red arrow $def$ to make the object $\lambda s_{h_1}$ and $s_{\lambda h_1}$ becoming equivalent, i.e., there is a bijection between them. The broken line arrow represents the universal morphism to the direct sum object $\lambda s_{h_1}$.

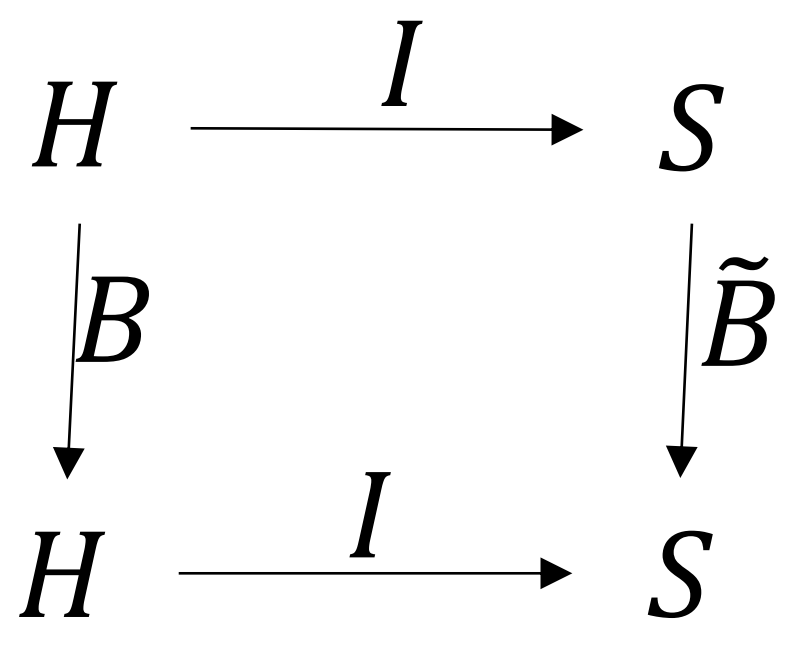

**Figure S3. The commutive diagram of $\boldsymbol{B}$ and $\boldsymbol{\tilde{B}}$.** Considering $I$ as the functor between interpreter space $H$ and cognitive space $S$, let $B$ represent an automorphism on $H$ and $\tilde{B}$ be the automorphism on $S$ so that the diagram above commutes.

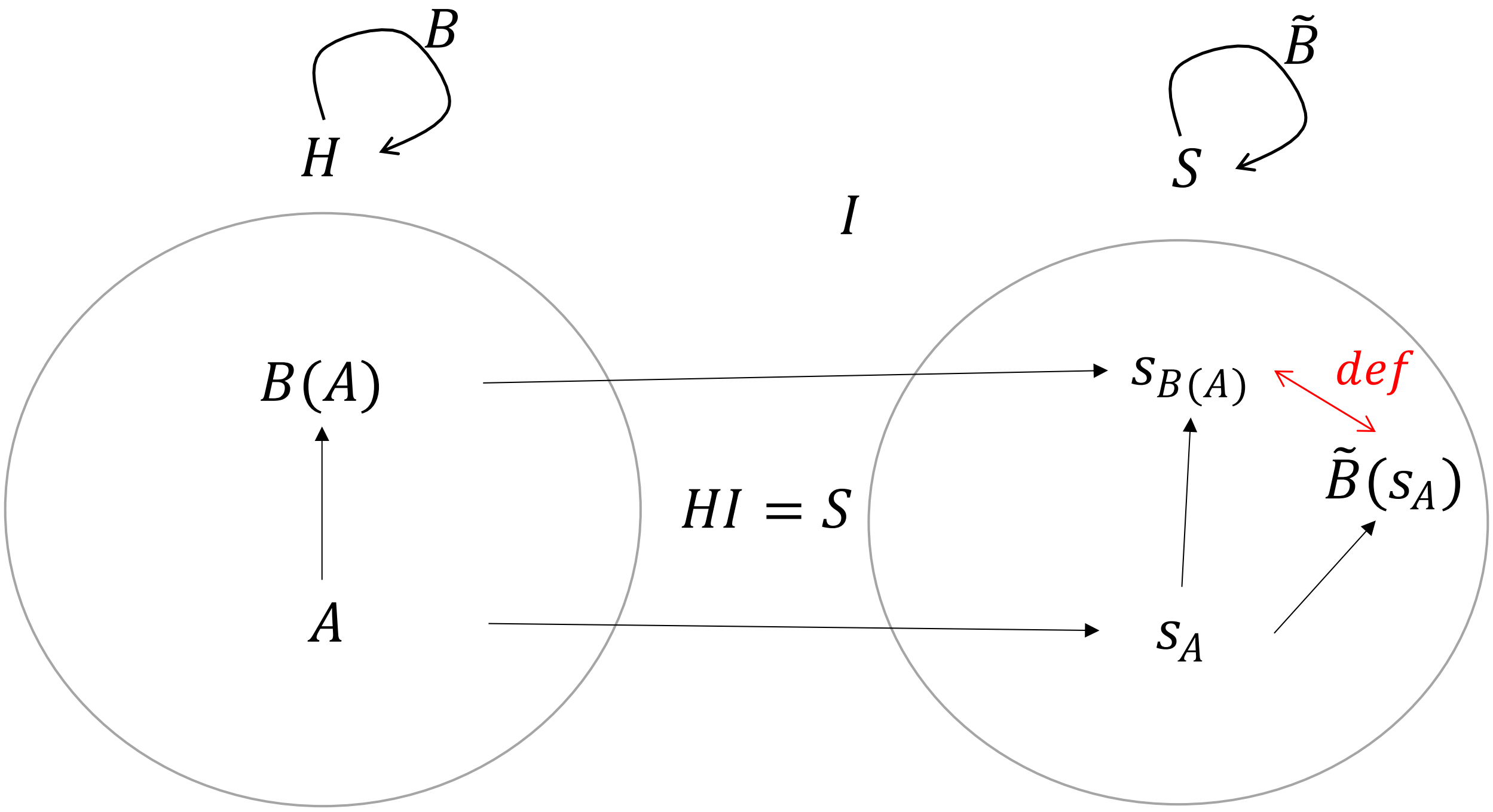

**Figure S4. the category diagram of the definition of the operator on operator.** With an automorphism $B$ on $H$, an object $A$ in $H$ is mapped into $B(A)$. With the functor $I$, the $A$ is mapped into $s(A)$ and $B(A)$ is mapped into $s_{B(A)}$. We define the corresponding automorphism $\tilde{B}$ on $S$ by giving the bijection $def$ between $s_{B(A)}$ and $\tilde{B}(s_A)$ so that the diagram in Fig. S3 commutes.

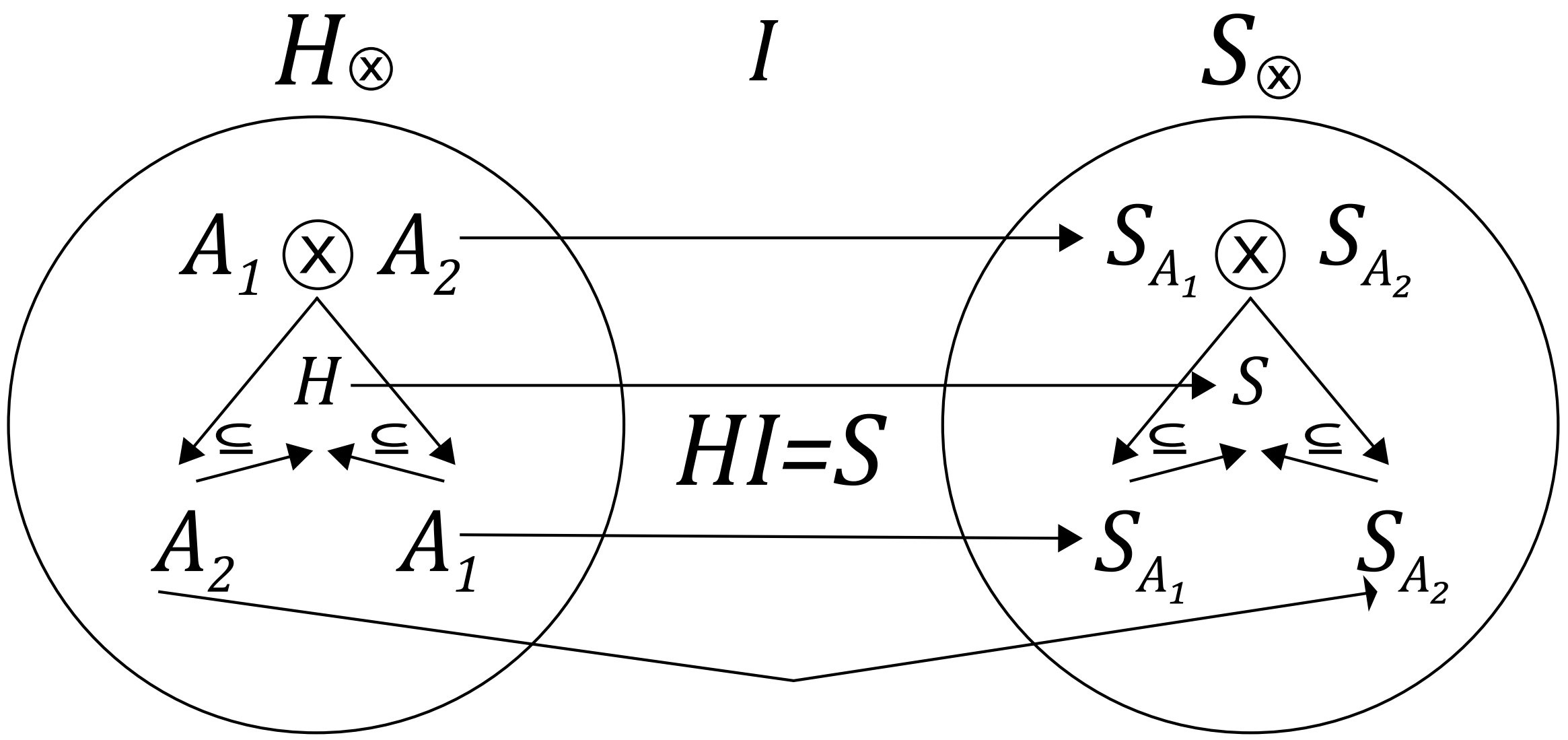

**Figure S5. the category diagram of the definition of the tensor space $H_\otimes$ and $S_\otimes$.** Here, the category $H_\otimes$ is the tensor space of interpreter space $H$ whose objects are the tensor products in linear algebra or the Descartes product in the category theory of the subspace of $H$. $S_\otimes$ is the tensor space of $S$ whose objects are the tensor products in linear algebra or the Descartes product in the category theory of the subspace of $S$. This figure shows that $A_1, A_2$ are the subspace of $H$ and their tensor product is in $H_\otimes$, noticing that $A_1, A_2$ and $H$ are in $H_\otimes$ as well since terminal object $1 \in H$ and for all $A \sqsubseteq H$, $A \otimes \{1\} \cong A$. Functor $I$ preserves the structure in $H_\otimes$ into $S_\otimes$ so that the diagram above commutes.

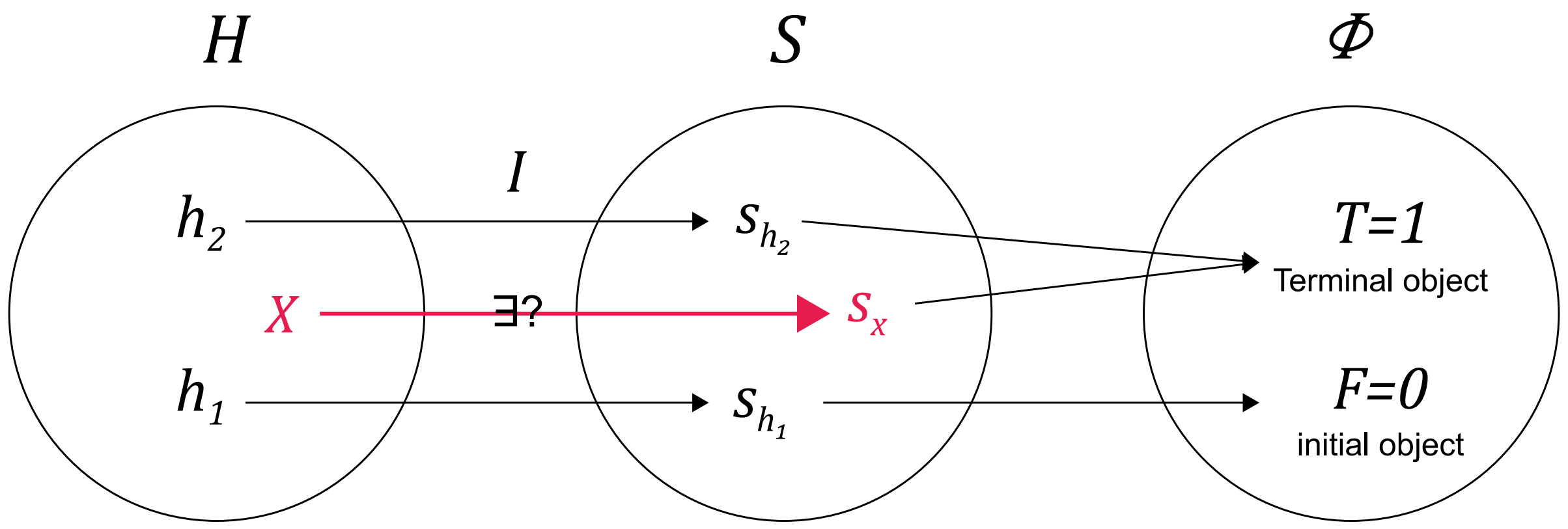

**Figure S6. the category diagram of the cognitive process**. Here $H$ is the interpreter space whose object is the interpreter such as $h_1$, $h_2$. The corresponding interpreted sentences or proposition or distribution such as $s_1, s_2$ are in the cognitive formal space $S$. We give each $s \in S$ a true or false value by the terminal object and the initial object in category Φ.

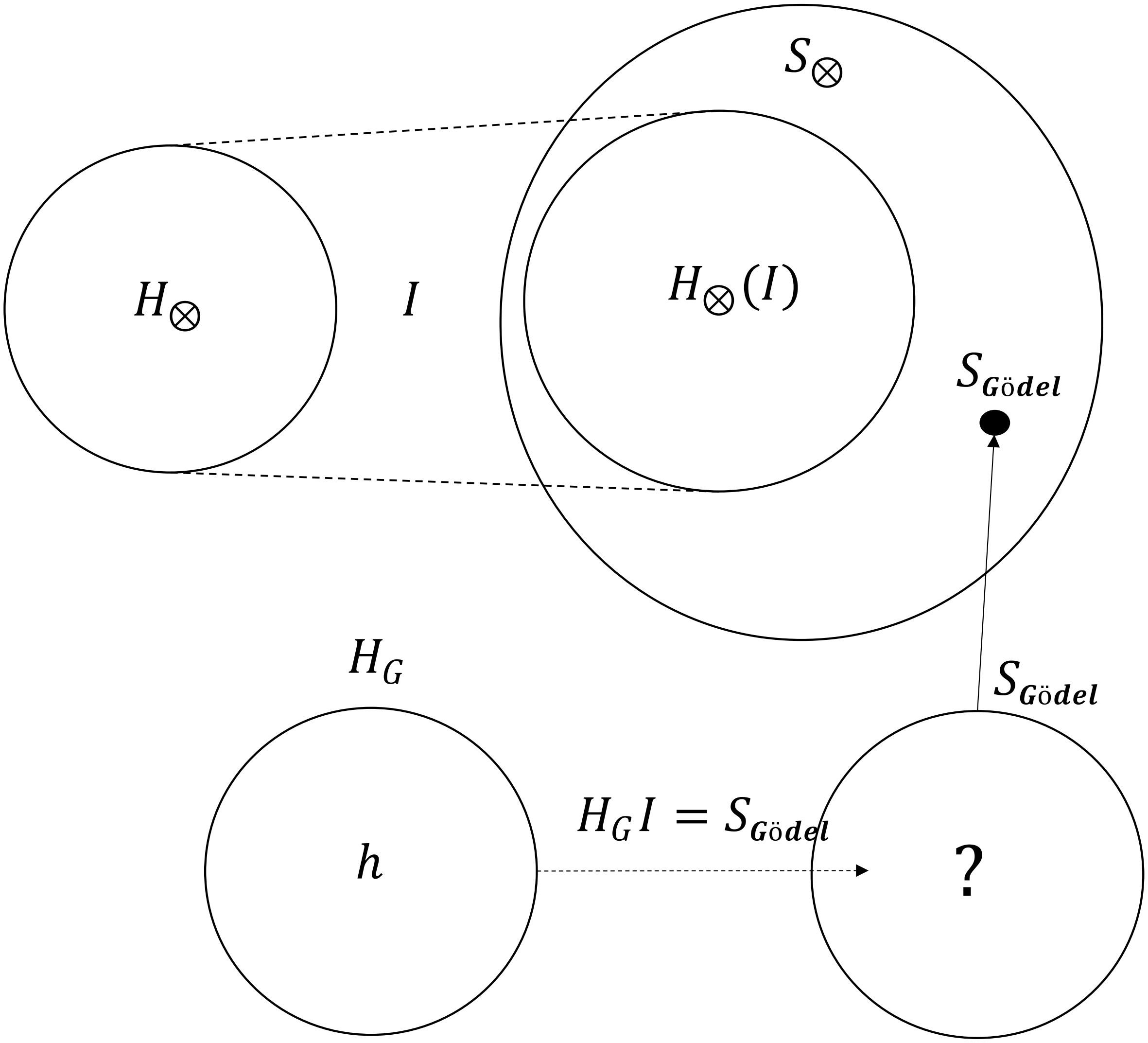

**Figure S7. the category diagram of the relationship between the tensor spaces and the incompleteness space.** The upper half of this figure shows that $H_{\otimes}$ mapped into $S_{\otimes}$ by $I$ could only get a part of $S_{\otimes}$ which is $H_{\otimes}(I)$. The other space of $S_{\otimes}$ is the $\boldsymbol{S}_{\boldsymbol{G}\text{ö}\boldsymbol{del}}$, whose dimension is $2^n - n - 1$ for n-dimension $H$. Here we discuss further in the lower half of this figure to find what kind of interpreter space $H_G$ so that $H_G I = \boldsymbol{S}_{\boldsymbol{G}\text{ö}\boldsymbol{del}}$. This discussion is equivalent to finding the more elaborate structure of incompleteness space $\boldsymbol{S}_{\boldsymbol{G}\text{ö}\boldsymbol{del}}$.

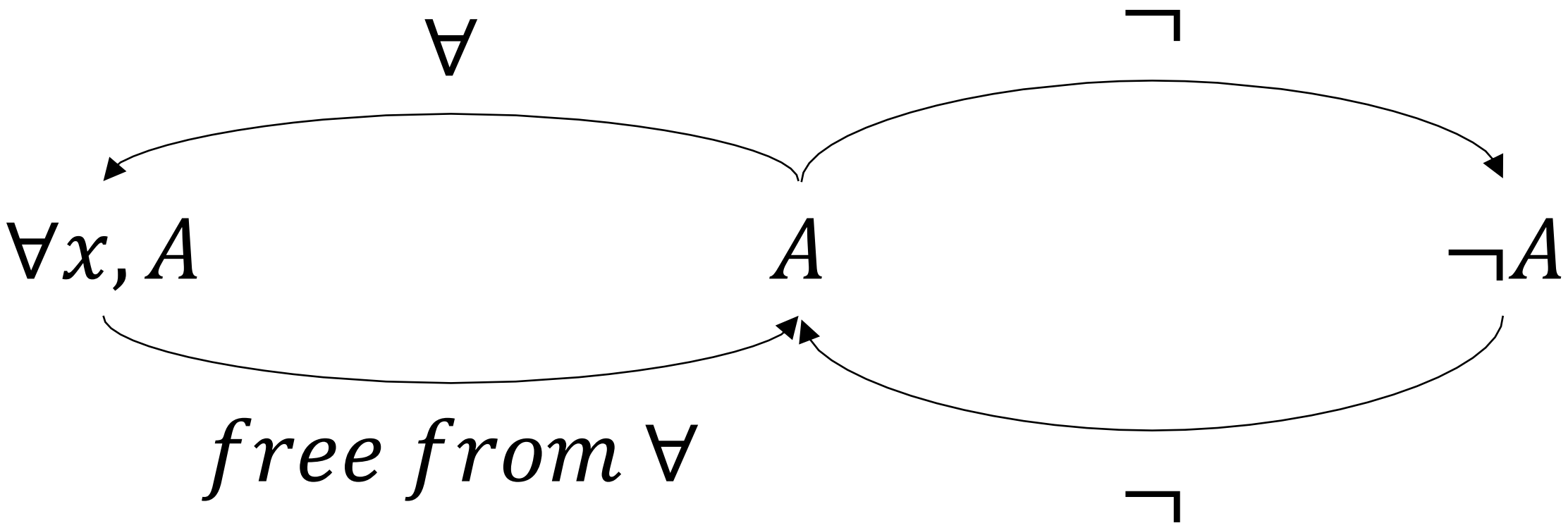

**Figure S8. the diagram of the inequivalent loop.** For $A$ is a proposition in formal system, $\neg$ is the logical negation operator, and $\neg A$ is the negative proposition of $A$. $\forall$ is 'for all', and $x$ is a free variable from $A$, i.e., variable $x$ is not in $A$. Therefore, one could construct different independent proposition by adding negation operator or the universal quantifier. Hence, the two inequivalent loops are existing so that we could apply our general incompleteness theorem on the first-order predicate logic.

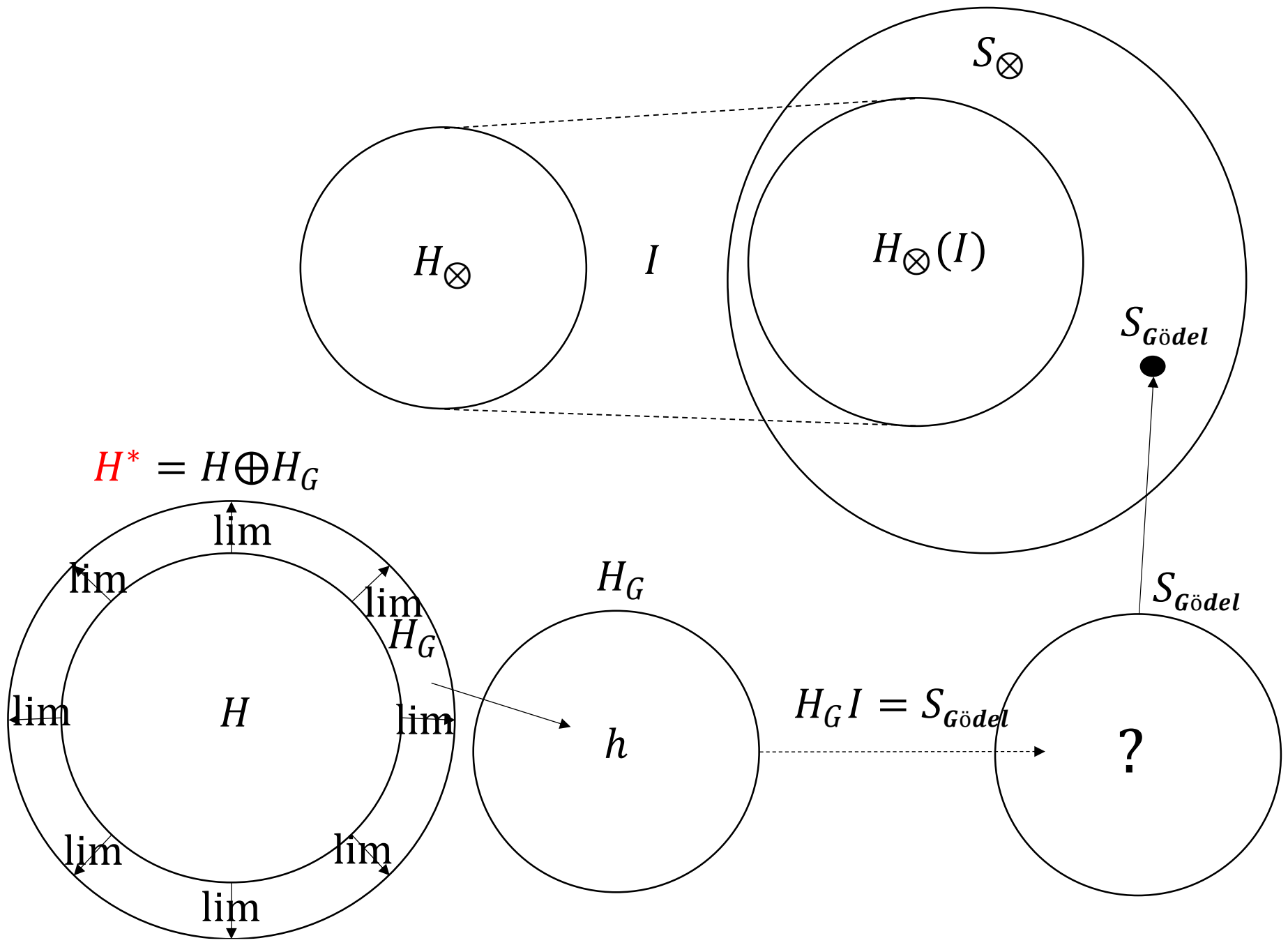

**Figure S9. the category diagram of the mapping structure of the incompleteness space.** The $H^*$ describes the limitations of the cognitive capability of the interpreter space $H$. Other notations are the same as in Fig. S7. We give the specific way to construct the incomplete interpreter space, the PAC-learnable interpreter space but cannot be found in $H$. This way is the same as the construction approach in section 4.2, which is exactly the Cantor diagonal construction.

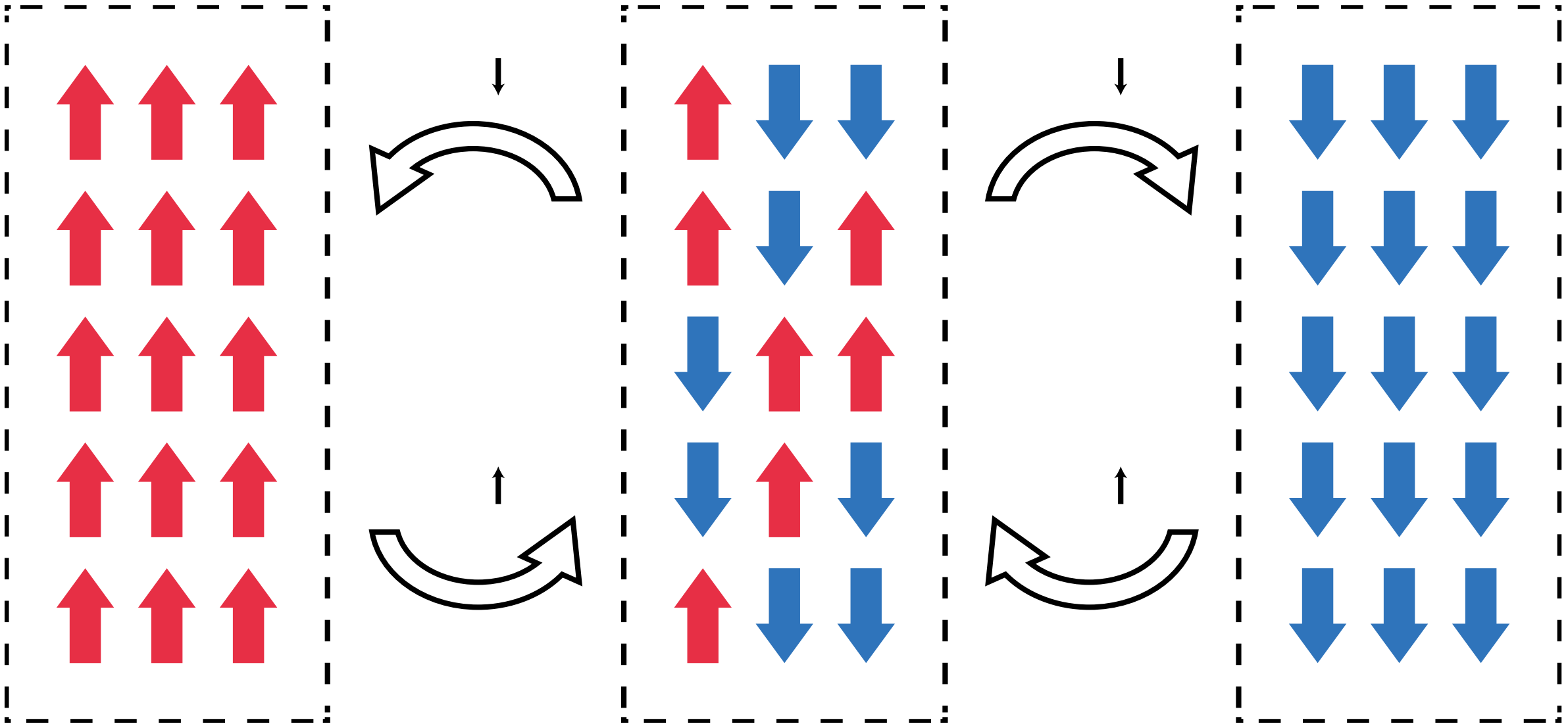

**Figure S10. the diagram of the loop structure of the ferromagnetic phase transformation.** The up and down arrows describe the spin as in the Ising model. $T$ means the temperature, $T\uparrow$ means raising the temperature and $T\downarrow$ means decreasing the temperature. The left and right diagram represent the ferromagnetic phase. The diagram in the middle represents the paramagnetic phase.

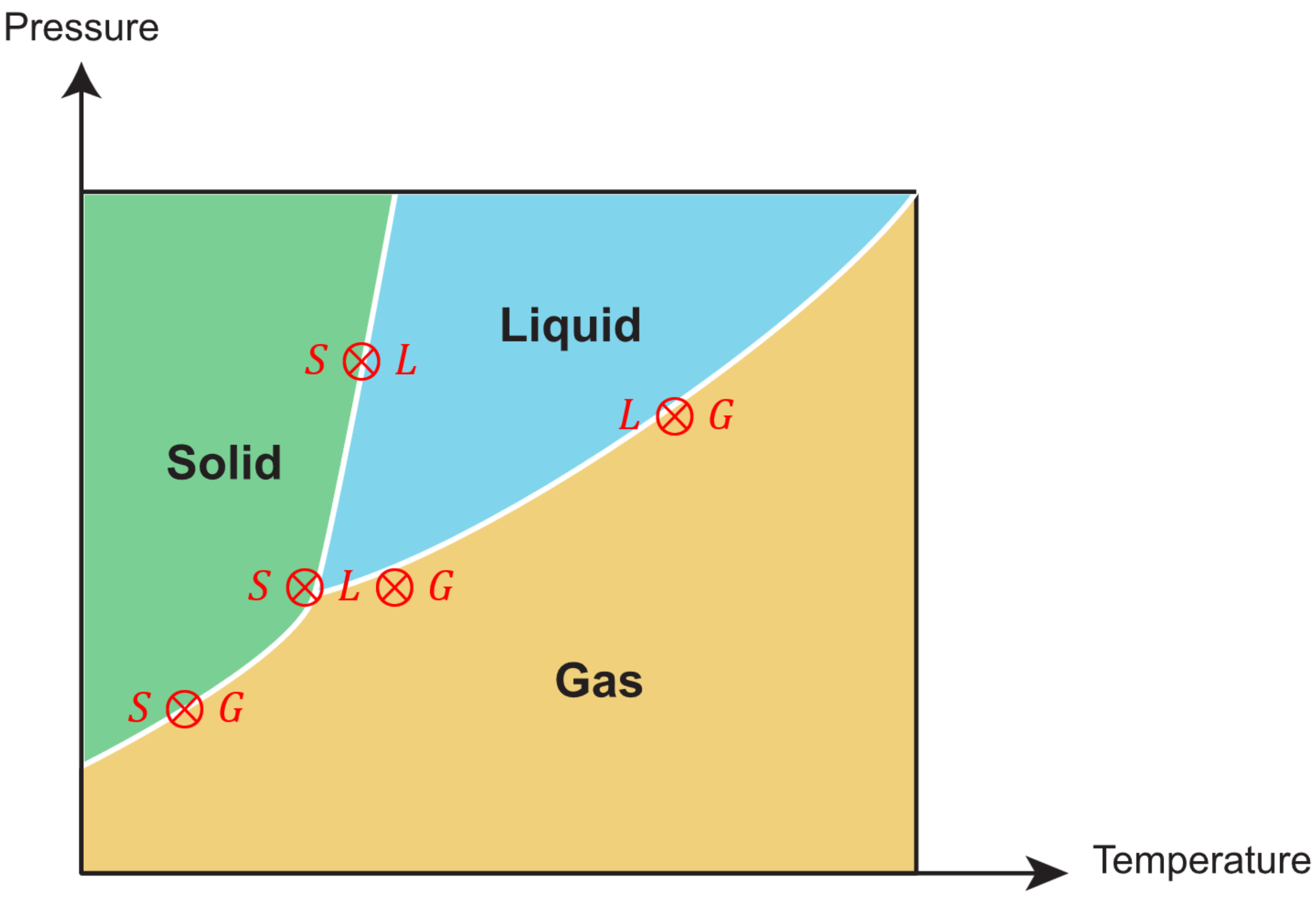

**Figure S11. the diagram of the three-phase transition of water.** The tensor products on the edges capture the phase of coexistence of multiple states. $S \otimes G$ is corresponding to solid-gas coexistence. $S \otimes L$ is corresponding to solid-liquid coexistence. $L \otimes G$ is corresponding to liquid-gas coexistence. $S \otimes L \otimes G$ is corresponding to the triple point of water.

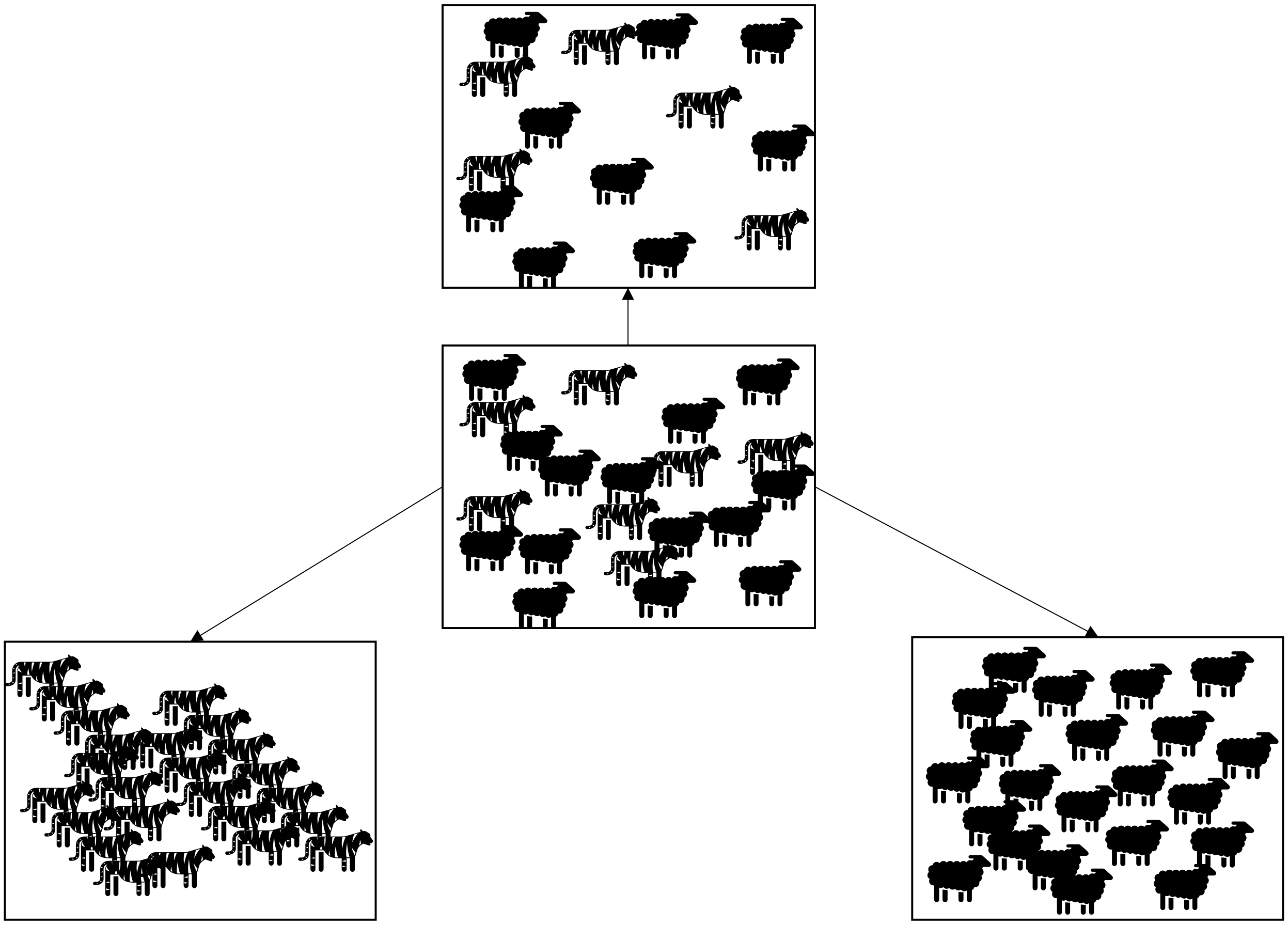

**Figure S12. the diagram of the predator-prey model (Lotka-Volterra model).** The arrows capture the three different kinds of symmetry breaking. The upward arrow indicates that predators and prey enter a state of steady fluctuations in population size. The arrow at the lower left corner indicates that it has entered a state where the predator is absolutely dominant and the prey has become extinct. The arrow at the lower right corner indicates that it has entered a state where the prey is absolutely dominant and the predator is extinct. The three different arrows are corresponding to three different kinds of symmetry breaking so that we can describe the critical state in the center with incompleteness space.